\documentclass[fleqn,usenatbib]{mnras}

\usepackage{amssymb}	

\usepackage{newtxtext,newtxmath}

\usepackage[T1]{fontenc}
\usepackage{ae,aecompl}

\DeclareRobustCommand{\VAN}[3]{#2}
\let\VANthebibliography\thebibliography
\def\thebibliography{\DeclareRobustCommand{\VAN}[3]{##3}\VANthebibliography}

\usepackage{gensymb}

\usepackage{graphicx}	
\usepackage{subcaption}
\usepackage{amsmath}	
\usepackage{ulem}
\usepackage{color}     
\usepackage{svg}
\usepackage{enumitem}
\renewcommand{\vec}[1]{\mathbf{#1}}

\newcommand{\DS}{\displaystyle}
\newcommand{\pd}[2]{\frac{\partial #1}{\partial #2} }
\newcommand{\HALF}{\frac{1}{2}}

\newcommand{\av}[1]{\left< {#1} \right>}
\newcommand{\quotes}[1]{``#1''}
\newcommand{\tens}[1]{\mathsf{#1}}

\title[3D relativistic magnetic reconnection]{2D or not 2D? Exploring 3D relativistic magnetic reconnection dynamics with highly accurate numerical simulations}

\author[Berta et al.]{
V. Berta$^{1,2,3},$\thanks{E-mail: vittoria.berta@unito.it}
M. Bugli$^{4,5,3}$,
A. Mignone$^{3,6}$,
G. Mattia$^{7}$,
L. Del Zanna$^{8,9}$, and
S. Truzzi$^{3}$
\\
$^{1}$Canadian Institute for Theoretical Astrophysics, University of Toronto, Toronto, ON M5S 3H8, Canada\\ 
$^{2}$Department of Astronomy, Columbia University, 550 W 120th St, New York, NY 10027, USA\\
$^{3}$Dipartimento di Fisica, Università degli Studi di Torino , Via Pietro Giuria 1, I-10125 Torino, Italy\\
$^{4}$Institut d’Astrophysique de Paris, UMR 7095, CNRS \& Sorbonne Universit\'e, 98 bis bd Arago, 75014 Paris, France\\
$^{5}$Université Paris-Saclay, Université Paris Cité, CEA, CNRS, AIM, F-91191 Gif-sur-Yvette, France\\
$^{6}$INFN - sezione di Torino , Via Pietro Giuria 1, I-10125 Torino, Italy\\
$^{7}$Max-Planck-Institut für Astronomie, Königstuhl 17, 69117 Heidelberg, Germany  \\
$^{8}$Dipartimento di Fisica e Astronomia, Università degli Studi di Firenze, L.go E. Fermi 2, I-50125 Firenze, Italy \\
$^{9}$INAF - Osservatorio Astrofisico di Arcetri, L.go E. Fermi 5, I-50125 Firenze, Italy \\
}
\date{Accepted XXX. Received YYY; in original form ZZZ}

\pubyear{2023}

\begin{document}

\maketitle


\begin{abstract}

Fast reconnection in magnetically dominated plasmas is widely invoked in models of dissipation in pulsar winds, gamma-ray flares in the Crab nebula, and to explain the radio nanoshots of pulsars.
When current sheets evolve reaching a critical inverse aspect ratio, scaling as $S^{-1/3}$ with the plasma Lundquist number, the so-called \textit{ideal} tearing instability sets in, with modes growing, independently of $S$, extremely rapidly on timescales of only a few light-crossing times of the sheet length.
We present the first set of fully 3D simulations of current-sheet disruption triggered by the ideal tearing instability within the resistive relativistic MHD approximation, as appropriate in situations where the Alfv\'en velocity approaches the speed of light.
We compare 3D setups with different initial conditions with their 2D counterparts, and we assess the impact of dimensionality and of the magnetic field topology on the onset, evolution, and efficiency of reconnection. 
In force-free configurations, 3D runs develop ideal tearing, secondary instabilities, and a thick, turbulent current layer, sustaining dissipation of magnetic energy longer than in 2D.
In pressure-balanced current sheets with a null guide field, 2D reference runs show the familiar reconnection dynamics, whereas in 3D tearing dynamics is quenched after the linear phase, as pressure-driven modes growing on forming plasmoids outcompete plasmoid coalescence and suppress fast dissipation of magnetic energy.
Taken together, these results suggest that the evolution and efficiency of reconnection depend sensitively on the local plasma conditions and current-sheet configuration, and can be properly captured only in fully 3D simulations.

\end{abstract}

\begin{keywords}
magnetic reconnection - methods: numerical - (magnetohydrodynamics) MHD - plasma - relativistic processes – three dimensional
\end{keywords}


\section{Introduction}
%
%

Astrophysical high-energy sources consistently exhibit non-thermal photon spectra, implying the presence of mechanisms that accelerate particles to power-law energy distributions \citep[e.g.,][]{Giannios_2005, Celotti_Ghisellini_2008, Zhang_etal_2020}. 
Such sources include flares from pulsars' and black holes' magnetospheres \citep[e.g.,][]{Lyubarskii_1996, Baty_2013, Uzdensky_Spitkovsky_2014, Cerutti_2014, Olmi_PWN_2016, Philippov_Spitkovsky_2018, Nathanail_etal_2022, Ripperda_etal_2022}, fast radio or gamma-ray bursts \citep[e.g.,][]{Zhang_Yan_2011, McKinney_Uzdensky_2012, Beniamini_Giannios_2017, Mahlmann_etal_2022, Most_Philippov_2023}, or in the coronae of black holes’ accretion disks at the base of the relativistic jets of Active Galactic Nuclei and microquasars \citep[e.g.,][]{Sironi_Petropoulou_Giannios_2015, Petropoulou_etal_2016}.
To explain the electromagnetic emission we observe, the aforementioned mechanisms must operate efficiently in the relativistic regime, where the average magnetic energy per particle exceeds the rest mass energy \citep[e.g.,][]{Lyutikov_Uzdensky_2003}.
At macroscopic scales (i.e., scales orders of magnitude larger than the skin depth of the plasma), the dynamics of astrophysical flows is well described within the ideal Magnetohydrodynamics (MHD) approximation, under which the Alfv\'en theorem (i.e., flux freezing) applies.
However, as the flow evolves, localized regions with strong gradients and intense currents may develop. 
In these regions, non-ideal effects become significant and can no longer be neglected in the energy-momentum balance \citep[e.g.,][]{Qian_rHARM_2017, Vourellis_2019, Ripperda_2019a, Ripperda_Bacchini_Philippov_2020, Mattia2023, Nathanail_2025}.

In this context, magnetic reconnection has gained popularity over the past few decades as an efficient and universal mechanism that can power the transient, burst-like emissions observed \citep[e.g.,][and references therein]{Sironi_2025_RecReview}. 
It occurs when magnetic field lines of inverse polarity annihilate, rapidly releasing the energy stored in the magnetic field as heat, fast plasma flows, and energetic particles \citep[e.g.,][]{Zenitani_2001, Sironi_Spitkovsky_2014, Werner_2016, Barkov_Komissarov_o2016, Sironi_2022, Zhang_2023}.
For this mechanism to operate, the plasma must depart from ideal behaviour, so that the Alfv\'en theorem no longer applies locally and magnetic field lines are free to change their topology.
In the resistive MHD approximation, the process is regulated by the induction equation where two timescales can be identified: the ideal (alfv\'enic) one $\tau_A = L/c_A$, associated with dynamical variations of the magnetic field, and the diffusive one $\tau_D = L^2/\eta$, associated with the diffusion of the magnetic field. 
Here $L$ is a characteristic scale of variation of the equilibrium magnetic field, $c_A$ is the Alfv\'en velocity, and $\eta$ is the local magnetic resistivity. 

Classical steady-state reconnection models - such as the Sweet-Parker \citep{Sweet_1958, Parker_1957} - fail to predict the highly energetic transient events observed in astrophysical plasmas. 
These models have therefore been supplanted by descriptions of bursty, tearing-unstable reconnection regions, characterized by the continuous formation and expulsion of multiple secondary islands \citep[i.e., plasmoids][]{Zelenyi_Krasnoselskikh_1979, Bhattacharjee_2009, Uzdensky_etal_2010, Loureiro_Uzdensky_2016}.
The tendency of current sheets to fragment into chains of magnetic islands or plasmoids was originally formalized by \citet{Furth_etal_1963}. 
The predicted reconnection timescale, $\tau_{\mathrm{rec}} \propto \sqrt{\tau_D \tau_A}$, implies a positive correlation with the Lundquist number, $S = \tau_D / \tau_A = L c_A / \eta$, which, however, is insufficient to account for the explosive flaring activity observed, given the extremely large values of $S$ typical of relativistic astrophysical plasmas.

\citet{Pucci_Velli_2014} presented a theoretical model leading to fast reconnection regimes in the single-fluid resistive MHD approximation. 
Their work showed that, providing large enough values of the Lundquist number ($S \geq 10^6$), Sweet-Parker current sheets are unlikely to form in astrophysical plasmas. 
Instead, when thin current sheets of inverse aspect ratio $a/L$, where $a$ is the current sheet half width, and $L$ is the current sheet half length, scale as $S^{-1/3}$, a critical transition occurs: the growth rate of the tearing mode instability becomes asymptotically independent of $S$, with modes growing on ideal time scales $\propto \tau_A$.
The ``ideal'' tearing instability was verified numerically for the first time by \citet{Landi_etal2015}, who analyzed the stability of a current sheet undergoing ideal tearing through two-dimensional (2D) compressible MHD simulations, and by \citet{Tenerani_2015}, who confirmed that in collapsing current sheets, fast reconnection is triggered before a Sweet-Parker-type configuration can form. 
The fast reconnection regime was also extended to account for visco-resistive \citep{Tenerani_visco_2015}, Hall effects \citep{Pucci_Hall_2017, Papini_etal2019}, or in double-current sheet systems \citep{Komissarov_etal_2007, Baty_2017_ITMI}, always with 2D numerical probes.
\citet{DelZanna_2016} extended this framework to resistive relativistic MHD, using 2D numerical simulations.
To date, studies of reconnection dynamics driven by the ideal tearing mode instability within fully three-dimensional (3D) resistive spacial-relativistic MHD(ResRMHD) frameworks remain unexplored. 

Investigating 3D relativistic reconnection under different configurations is crucial to overcome the intrinsic limitations of two-dimensional models and to assess the extent to which such models capture the dynamics occurring in nature.
In this paper, we present simulations of 3D relativistic magnetic reconnection triggered by the ideal tearing mode instability by solving the equations of resistive relativistic MHD \citep{Komissarov_2007, DelZanna_2016, Mignone_etal_2019, Ripperda_2019a}. 
While 2D reconnection can be bursty due to the stochastic formation, motion, and merging of plasmoids, these events generally follow a predictable evolution toward a unique final state \citep[e.g.,][]{Shibata_Tanuma_2001, Loureiro_2007, Komissarov_etal_2007, Huang_Bhattacharjee_2013, Guo_2016, Hakobyan_2021}. 
In contrast, 3D reconnection exhibits greater variability, with stochastic processes generating a thicker turbulent layer in diverse ways \citep{Cerutti_2014}. 
Following \citet{Werner_Uzdensky_2021}, we emphasize that reconnection should be understood as the conversion of unreconnected to reconnected magnetic flux, conserving the total flux. 
In 2D, this conservation is straightforward, but in 3D it is challenging to define and measure it precisely. 
Rather than adopting a strict definition of reconnection, we focus on assessing the properties of the current layer, such as magnetic and plasma energy, and the layer dynamics. 
Reconnection in 3D therefore refers broadly to any magnetic-energy-depleting evolution—including flux annihilation—of thin current sheets that could, in principle, undergo 2D-like reconnection, while references to reconnecting flux specifically denote the 2D flux-conserving process.

Resistive relativistic MHD provides a cost-effective framework to capture the impact of magnetic dissipation—occurring on microscopic scales—within the macroscopic evolution of relativistic astrophysical plasmas \citep[e.g.][]{Zanotti_Dumbser_2011, Ripperda_2019b, Bransgrove_Ripperda_2021, Mattia_etal_2024, Grehan_2025}. 
At the same time, the role of physical and numerical resistivity in ResRMHD simulations is also a key question to disentangle \citep{Mattia2023, Grehan_2025, Bugli_etal_2024, Das_2025, Ripperda_2026}.
A sufficiently high grid resolution is required to ensure that dissipation is set by the imposed physical resistivity rather than by numerical diffusion. 
This is particularly important for capturing ideal tearing, which develops in low-resistivity regimes and demands sufficient resolution to discretize the smaller dissipation scales at reconnection onset \citep{Puzzoni_2021}. 

In this context, recent advances in $4^{\rm th}$-order accurate finite-volume schemes \citep{Verma_etal2019, Berta_etal_2024, Mignone_Berta_2024} offer significant improvements over traditional $2^{\rm nd}$-order methods. 
Their lower numerical dissipation enables exploration of reconnection processes with reduced grid resolution requirements while retaining stability and computational efficiency \citep[see, e.g., \S 4.5 of][]{Mignone_Berta_2024}.
These capabilities, combined with the power of Graphics Processing Units (GPUs) to reduce computational demands help making fully 3D ResRMHD simulations more accessible.
In \S \ref{sec:equations} we describe the working framework and the simulation setups. 
In \S \ref{sec:results} we describe the results of the simulations, discussing the results obtained in \S \ref{sec:summary}.

\section{Method description}
\label{sec:equations}
%

In this section, we describe the system of equations for resistive relativistic MHD in a flat Minkowskian spacetime, as well as the numerical setup and the initial conditions assumed in our simulations.

\subsection{Equations of ResRMHD}

The system of ResRMHD equations consists of a set of conservation laws in the form of hyperbolic differential equations with non-zero source terms \citep{Komissarov_2007, Mignone_etal_2019}
\begin{equation}
  \pd{U}{t} = -\nabla\cdot\tens{F}(U) + S_e(U) + S_i(U),
\end{equation}
where $U$ is the vector of conserved variables, $F$ is the array of corresponding fluxes, and $S_e$ and $S_i$ are, respectively, the non-stiff and stiff contributions to the source term, handled either explicitly or implicitly in our numerical method.
The explicit form of such equations for the different conserved variables yields
\begin{equation}
\begin{array}{l}
\DS\pd{D}{t} + \nabla\cdot(D\vec{v}) = 0 , \\ \noalign{\medskip}
\DS\pd{\vec{m}}{t} + \nabla\cdot [\rho h \gamma^2\vec{v}\vec{v} - \vec{E}\vec{E} - \vec{B}\vec{B} + (p + \frac{E^2 + B^2}{2})\,\tens{I}] = 0 , \\ \noalign{\medskip}
\DS\pd{\cal E}{t} + \nabla\cdot\vec{m} = 0, \\ \noalign{\medskip}
\DS\pd{\vec{B}}{t} + \nabla\times\vec{E} = 0 , \\ \noalign{\medskip}
\DS\pd{\vec{E}}{t} - \nabla\times\vec{B} = -\vec{J}, 
\end{array}
\end{equation}
with the addition of the two non-evolutionary constraints
\begin{equation}
\begin{array}{l}
\nabla\cdot\vec{B} = 0, \\ \noalign{\medskip}
\nabla\cdot\vec{E} = q,
\end{array}
\end{equation}
where the speed of light has been set to $c = 1$, and the magnetic permeability $1/\sqrt{4\pi}$ has been absorbed into the definitions of the electromagnetic fields ${\vec B}$, and ${\vec E}$ (those measured in the fixed laboratory frame), as in Heaviside-Lorentz units.
The conserved fluid variables are
\begin{equation}
\begin{array}{lcl}
D & = & \rho\gamma,  \\  \noalign{\medskip}
\vec{m} & = & \rho h\gamma^2\vec{v} + \vec{E}\times\vec{B}, \\ \noalign{\medskip}
{\cal E} & = & \rho h\gamma^2 - p + \DS\frac{E^2 + B^2}{2},
\end{array}
\end{equation}
respectively the mass density, momentum density, and total energy density, measured in the laboratory frame. 
The primitive fluid variables $(\rho,{\vec v}, p)$ are the rest-mass density, the velocity, and the thermal pressure of the fluid, $h$ is the relativistic specific enthalpy function, while $\gamma$ is the Lorentz factor of the flow. 
In our models, we assume a Taub equation of state \citep{Mignone_2007}
\begin{equation} \label{eq:taub_eos}
    h = \frac{5}{2}\left(\frac{p}{\rho}\right) + \sqrt{\frac{9}{4}\left(\frac{p}{\rho}\right)^2 + 1} \, ,
\end{equation}
The source term is the electric current density $\vec{J}$, defined in terms of the electromagnetic and velocity fields and the electric charge density $q$ as
\begin{equation}
\vec{J} = q\vec{v} + \eta^{-1}\gamma [\vec{E} + \vec{v}\times\vec{B} - (\vec{E}\cdot\vec{v})\vec{v}],
\end{equation}
obtained by assuming a simple Ohm's law with a scalar resistivity $\eta$ in the comoving frame. 
The term proportional to $\eta^{-1}$ is usually large, due to the extreme conductivity of astrophysical plasmas, it is numerically stiff and requires an implicit treatment \citep{Palenzuela_etal2009, Mignone_etal_2019, Tomei_etal_2020, Mattia2023, Mignone_Berta_2024}.
In this work we adopt the numerical scheme proposed by \citet{Mignone_Berta_2024} to discretize the governing equations.




\subsection{Initial conditions}

We consider initial conditions corresponding to a static plasma ($\vec{v} = 0$), with the equilibrium defined as 
\begin{equation} \label{eq:equil}
  \begin{aligned}
     \vec{B} &= B_0 \left[0, \tanh(x/a), \zeta \, \text{sech}(x/a)\right] \, , \\
     p &= p_0 + \frac{B_0^2}{2}(1 - \zeta^2) \, \text{sech}^2(x/a) \, ,  \\
     \rho &= \rho_0 \frac{p}{p_0}\, ,
  \end{aligned}
\end{equation}
where the subscript ``$_0$'' refers to upstream quantities (i.e., values asymptotically far from the current sheet).
Here $a$ is the current sheet half-width, $B_0$, $p_0$, and $\rho_0$ are the upstream magnetic field, pressure, and density. The value of $\rho_0$ is fixed to unity, while $B_0$ and $p_0$ are defined through the parameters $\sigma_0$, the plasma (cold) magnetization, and $\beta_0$, the plasma beta, respectively defined as 
\begin{equation}\label{eq:sigma_beta}
    \sigma_0 = \frac{B_0^2}{\rho_0} \, , \qquad
    \beta_0 = \frac{2p_0}{B_0^2} \, .
\end{equation}
The parameter $\zeta$, originally introduced by \citet{Landi_etal2015}, allows one to impose different kinds of initial equilibria: specifically, values of $0 < \zeta < 1$ interpolate between a purely pressure-balanced configuration for $\zeta = 0$, and a force-free equilibrium for $\zeta = 1$.
The choice $p\propto\rho$ provides an initial uniform temperature across the whole domain $\Theta=\Theta_0= p_0/\rho_0 = \sigma_0\beta_0/2$, as well as the specific upstream enthalpy $h_0$, to be retrieved from Eq. (\ref{eq:taub_eos}).
The upstream value of the relativistic Alfv\'en velocity may be written as a function of $\sigma_0$ and $\beta_0$ as
\begin{equation}
    c_A^2 = \frac{B_0^2}{\rho_0 h_0 + B_0^2} = \frac{1}{1+\frac{5}{4}\beta_0+\sqrt{\frac{9}{16}\beta_0^2+\frac{1}{\sigma_0^2}}}\, ,
\end{equation}
also defining the Alfv\'enic crossing time of the sheet's half length $L$ as $\tau_A = c_A / L$.
The global Lundquist number is
\begin{equation}\label{eq:lund}
    S = \frac{c_A L}{\eta} \, ,
\end{equation}
where $\eta$ is the magnetic resistivity, here assigned by actually choosing a value for $S$ and simply inverting Eq. (\ref{eq:lund}).
In our simulations, the equilibrium magnetic field in Eq. (\ref{eq:equil}) is prescribed by means of the vector potential as
\begin{equation} \label{eq:A_equil}
   \vec{A} = aB_0 \left[ 0, \zeta\rm{arctan\left(sinh(x/a)\right)}, -\rm{log\left(\cosh(x/a)\right)} \right] \, ,
\end{equation}
whereas the electric field is initially vanishing.

The magnetic field is perturbed using $N_m \times N_n = 10 \times 10$ modes, defined as  
\begin{align} \label{eq:force_free}
    \delta A_z = B_0 \frac{\epsilon}{N_m N_n} \sum_{m=1}^{N_m} \sum_{n=1}^{N_n} & \frac{1}{k_{y,m}} \sin(k_{y,m} y + \phi_m) \nonumber \\
    & \quad \sin(k_{z,n} z + \psi_n) \, \mathrm{sech}(x/a) \,.
\end{align}
The perturbation consists of a superposition of plane waves, where $\varepsilon$ is the initial perturbation amplitude, while $\phi_m$, and $\psi_n$ are randomly generated phases.
It is important to note that these phases are initialized to the same random values for all simulations presented here.
The wave-numbers discretized within the computational domain are
\begin{equation}\label{eq:kykz}
    k_{y,m} = 2\pi/\lambda_y = 2\pi m/L_y \; \text{and} \; 
    k_{z,n} = 2\pi/\lambda_z = 2\pi n/L_z \, ,
\end{equation}
with $L_y = L_z = 2L$. 
This configuration allows the growth of modes spanning from $k_{y,z} L = \pi$ to $k_{y,z} L = 10\pi$ exciting the current sheet. 

\subsection{Numerical setup}

Our setups are controlled by a set of six parameters, i.e., $\{\sigma_0, \beta_0, S, a/L, \varepsilon, \zeta \}$. 
The parameters $\sigma_0$, $\beta_0$, $S$, $a/L$, with $L = 1$, and $\varepsilon$ are set, respectively, to $\sigma_0 = 10$, $\beta_0 = 0.1$, $S = 10^6$, $a/L = S^{-1/3} = 10^{-2}$, such that the ideal tearing condition is fulfilled, and $\epsilon = 10^{-3}$, while $\zeta$ can be either 0 or 1.
This choice yields an initial upstream Alfv\'en velocity of $c_A \simeq 0.89$ (in units of $c$), and a resistivity of $\eta \simeq 0.89 \times 10^{-6}$.
The computational domain spans $x \in [-80a, 80a]$, $y, z \in [0, 2L]$, with reflective boundary conditions applied along $x$ and periodic boundaries along $y, z$.
Based on results provided in \citet{Mignone_Berta_2024}, simulations were run with a grid resolution of $N_{\rm grid} = \{1024 \times 768 \times 768\}$ for the 3D cases, and $N_{\rm grid} = \{1024 \times 768\}$ for the 2D.
The latter simulations have been conducted as reference cases by simply avoiding the dependence on $z$.
The final simulation time for all runs is $t = 60\tau_c$, with $\tau_c = c/L$ being the light-crossing time of the sheet's half length.
Table \ref{tab:models} summarizes the models taken under consideration in this article.

In order to minimize waves reflection in the $x$ direction while conserving the overall energy in the computational box, our domain is divided into a uniformly distributed, central region $x \in [-40a, 40a]$ where $75\%$ of the grid points are distributed, and by an outer region where the remaining $25\%$ of the points are disposed on a stretched grid.
This configuration achieves a reduction of $512$ grid points ($\simeq 33.3\%$) relative to a uniformly spaced grid over [-80a,80a] that would be required to maintain the same central resolution. 
The reduction translates directly into lower memory footprint and computational cost while retaining the target resolution where it is more required, i.e, closer to the current sheet.
In the buffer regions the scheme locally reverts to $2^{\rm nd}$-order accuracy in order to favor dissipation and waves damping.
\begin{table}
    \centering
    \caption{List of models with the corresponding initial parameters: equilibrium parameter $\zeta$, dimensions, and initial ratio of magnetic and thermal pressure on the current sheet with respect to the upstream.}
    \label{tab:models}
    \renewcommand{\arraystretch}{1.2}
    \begin{tabular}{c c c c c c}
        \hline
        \hline
        Model & $\zeta$ & Dimensions & $B_z(0)/B_0$ & $p(0)/p_0$\\
        \hline
        \texttt{ff2d}  & 1 & 2D & 1.0 & 1.0\\
        \texttt{pb2d}  & 0 & 2D & 0.0 & 11.0\\ 
        \texttt{ff3d}  & 1 & 3D & 1.0 & 1.0\\
        \texttt{pb3d}  & 0 & 3D & 0.0 & 11.0\\
        \hline
        \hline
    \end{tabular}
\end{table}

\section{Results}
\label{sec:results}
%
%

In this section, we present the dynamical evolution obtained from our 3D simulations. 
To assess the impact of dimensionality on the reconnection process, we also compare the properties of the 3D runs with those of corresponding 2D reference simulations. 
Our models are produced using the numerical scheme described by \citet{Mignone_Berta_2024}, employing the $5^{\rm th}$-order weighted essentially non-oscillatory (WENOZ) reconstruction of \citet{Borges_WENOZ2008}.
The scheme uses the IMEX SSP(4,3,3) Runge–Kutta time integration method of \citet{Pareschi_Russo2005}. 
The implicit portion is solved with an optimized Newton–Broyden multidimensional root finder, while the explicit portion computes edge-centered electromagnetic fields using the multidimensional five-wave Riemann solver of \citet{Mignone_etal_2019}, and updates zone-centered hydrodynamic variables using a Lax–Friedrichs Riemann solver. 
All simulations were carried out with the fully parallel, GPU-accelerated version of the \texttt{PLUTO} code \citep{Mignone_PLUTO2007, Rossazza_GPU_2025}.

\subsection{Force-free configuration}
\label{sec:FF}
%

\begin{figure}
  \centering
  \includegraphics[width=0.45\textwidth]{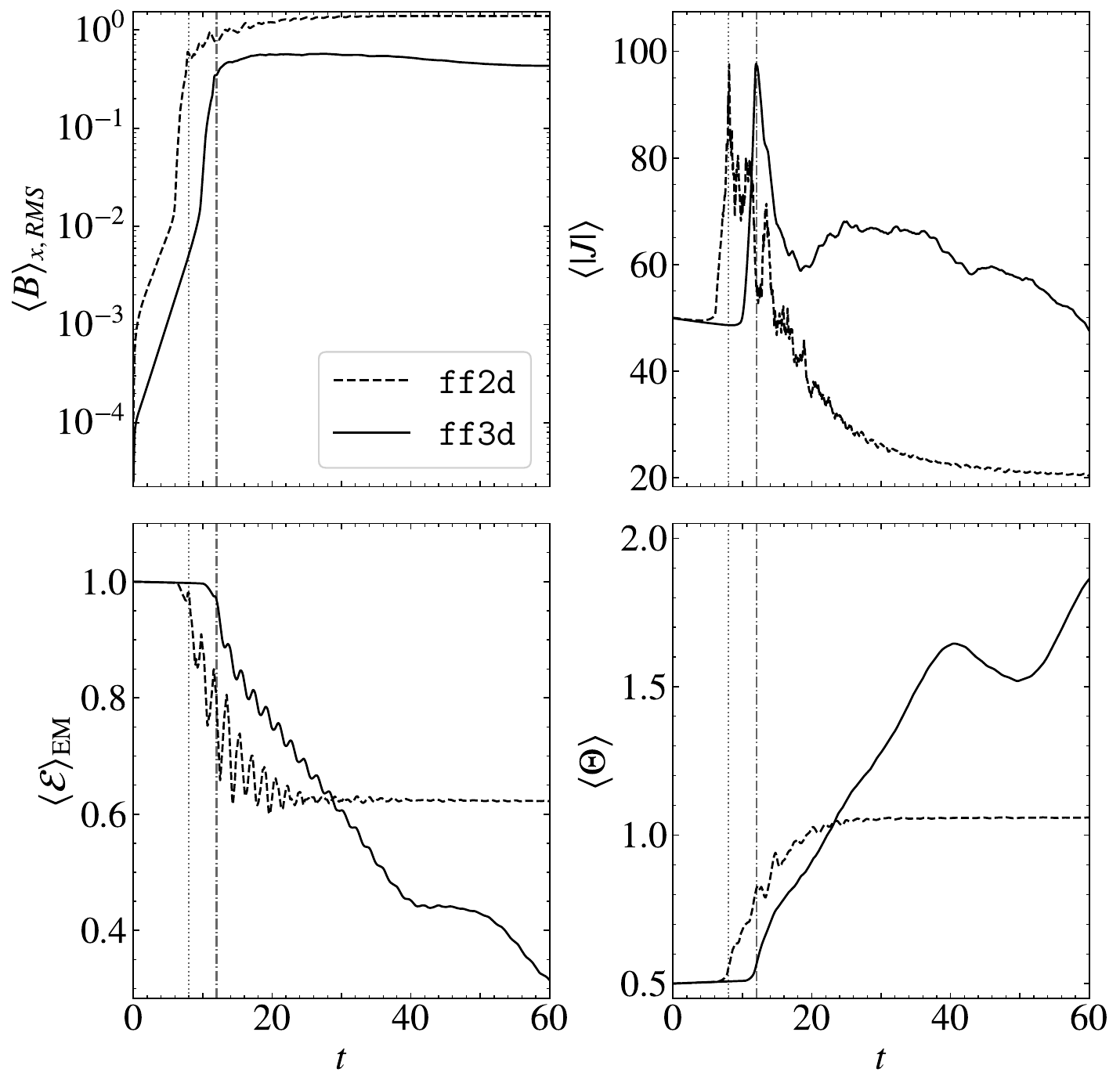}
  \caption{\small Temporal evolution of the root-mean-square of the magnetic field component $B_{x}$ (upper left panel), the magnitude of the total electric current $|J|$ (upper right), the electromagnetic energy ${\cal E}_{\text{EM}} = (E^2 + B^2)/(B_0^2)$ normalized to its initial value (bottom-left), and the temperature ${\Theta} = p/\rho$ (bottom-right) for simulations initialized with the force-free equilibrium.
  All volume-integrated quantities are indicated with the $\av{}$ symbol.
  Time is given in units of the light-crossing time of half the sheet length, $\tau_c = L/c$.
  The black dotted and dash-dotted vertical lines mark, respectively, the transition to the non-linear regime.
}
  \label{fig:FF_time_evolution}
\end{figure}

\begin{figure}
  \centering
  \includegraphics[width=0.45\textwidth]{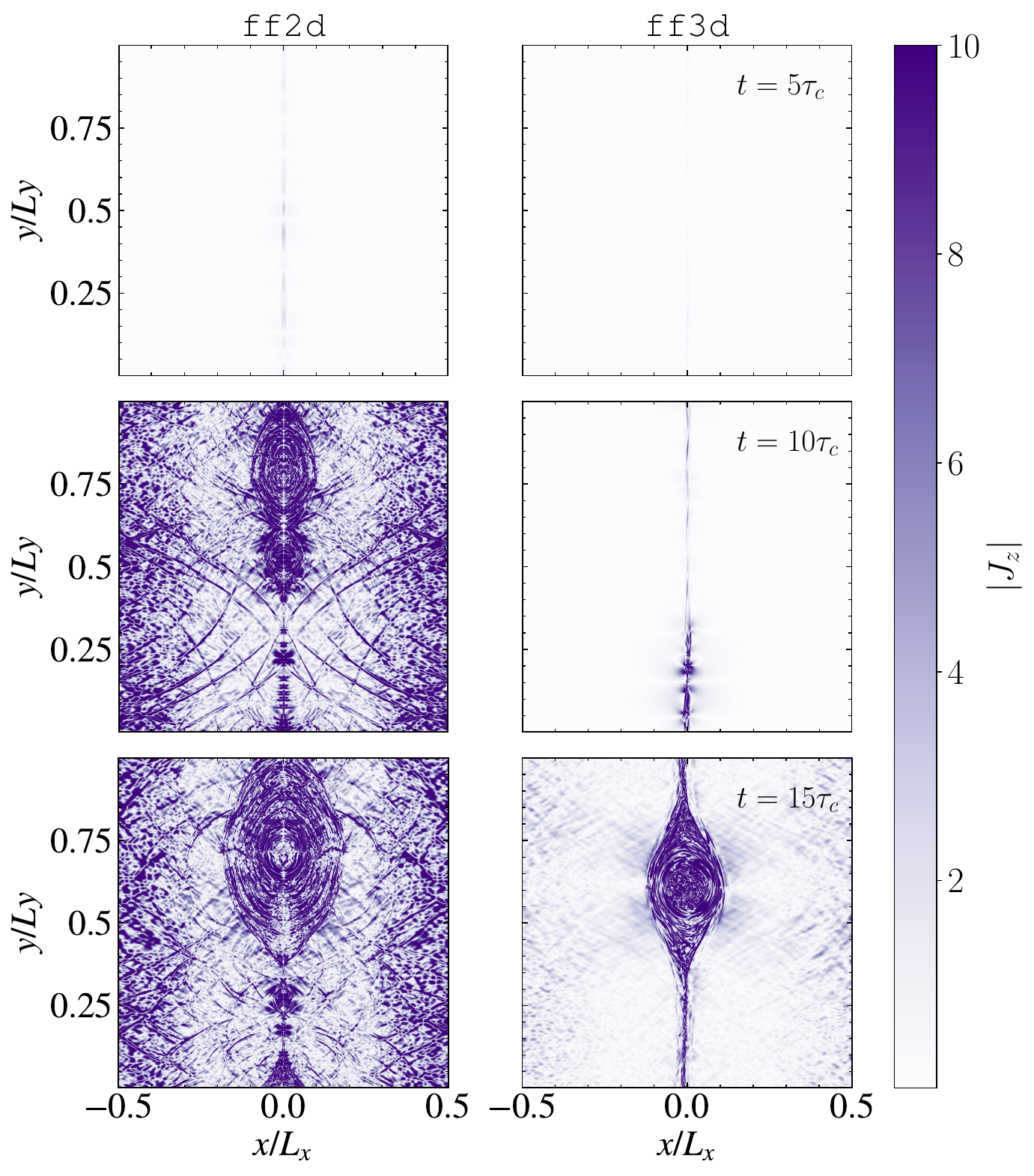}
  \caption{\small Snapshot of the module of the $z$-component of $|\nabla \times \vec{B}|$ used as a proxy for $|J_z|$ in the plane $z = 0$ for model \texttt{ff2d} (left column), and module \texttt{ff3d} (right column) at three different times: $t = 5\tau_c$ (top row), $t = 10\tau_c$ (middle row), $t = 15\tau_c$ (bottom row).   
}
  \label{fig:FF_xy_plane}
\end{figure}

Our analysis begins by comparing models \texttt{ff2d} and \texttt{ff3d}, as illustrated in Figures \ref{fig:FF_time_evolution} and \ref{fig:FF_xy_plane}.
%
%
%
%
The upper-left panel of Fig. \ref{fig:FF_time_evolution} shows the temporal evolution of the Root Mean Square (RMS) component of the reconnected magnetic field within the uniform region of the active computational domain:
\begin{equation} \label{eq:Brms}
        \av{B}_{x,\mathrm{RMS}} = \left(\frac{1}{80\pi^2 aL^2} \int_{-40a}^{40a} \int_0^{2L} \int_0^{2L} B_x^2(x,y,z,t) \, dx\, dy\, dz \right)^{1/2} \, .
\end{equation}
Alongside are plotted several other relevant volume-integrated quantities: the norm of $|\nabla \times \vec{B}|$ used as a proxy for the non-relativistic electric current $\av{|J|}$; the electromagnetic energy normalized by its initial value $\av{\mathcal{E}}_{\mathrm{EM}} = \left(\av{E^2} + \av{B^2}\right)/\av{B_0^2}$; and the average temperature $\av{\Theta} = \av{p/\rho}$.
%
%
%
%
Complementary spatial information is provided in Fig. \ref{fig:FF_xy_plane}, which shows snapshots of the magnitude of the $z$-component of $|\nabla \times \vec{B}|$ in the plane $z=0$. 
The left and right columns correspond to models \texttt{ff2d} and \texttt{ff3d}, respectively, and each row represents a different stage of the early evolution: $t = 5\tau_c$ (top), $t = 10\tau_c$ (middle), and $t = 15\tau_c$ (bottom).

At early times, the linear phase of the tearing instability shown by $\av{B}_{x,\mathrm{RMS}}$ in the upper-left panel of Fig. \ref{fig:FF_time_evolution} displays the characteristic exponential growth in both the \texttt{ff2d} and \texttt{ff3d} simulations, ending around $t = 8\tau_c$ for the 2D case (dotted vertical line). 
On the other hand, the 3D model evolves more slowly, owing to a redistribution of the energy imparted by the initial perturbation over three dimensions rather than two.
The resulting linear phase extends to $t = 12\tau_c $ (dash-dotted vertical line), attaining a lower value of $ \av{B}_{x,\mathrm{RMS}} $.
During the linear evolution, two distinct slopes are observed: the first is associated with the ideal tearing instability, whereas the second corresponds to the linear growth phase of plasmoid coalescence.
This is also evident from the top row of Fig. \ref{fig:FF_xy_plane}, where, at $t = 5\tau_c$, the growth of the tearing instability in 3D is delayed compared to the 2D case.
Towards the end of the linear phase the current sheet undergoes fragmentation and plasmoid formation.
This transition is visible both in the emergence of sharp peaks in the integrated current $\av{|J|}$ (upper-right panel of Fig. \ref{fig:FF_time_evolution}), and in the structures appearing in the middle and bottom rows of Fig. \ref{fig:FF_xy_plane}.
By $t = 10\tau_c$, the 2D simulation has entered the non-linear stage, exhibiting a highly structured pattern of current filaments and ripple-like features caused by the non-linear interaction of waves.
At the same time, the 3D model is still in the linear phase, with the first plasmoids forming within the central current layer.
By $t = 15\tau_c$, the 2D simulation has developed a large, symmetric plasmoid, whereas the 3D simulation produces a more irregular current structure with reduced symmetry and wave interactions, reflecting the additional degrees of freedom available in 3D.

The evolution of the energy diagnostics follow closely.
The total electromagnetic energy $\av{\mathcal{E}}_{\mathrm{EM}}$ (bottom-left panel of Fig. \ref{fig:FF_time_evolution}) remains nearly constant until the end of the corresponding linear phases, after which $\av{\mathcal{E}}_{\mathrm{EM}}$ begins to decrease as it is converted—primarily into heat—while the temperature $\av{\Theta}$ (bottom-right panel of Fig. \ref{fig:FF_time_evolution}) rises. 
Indeed, the rate of magnetic energy conversion during the early phase ($t \leq 30\tau_c$) is lower in 3D than in the 2D case.
Differences in the efficiency of magnetic energy release between the two models become even more significant after $t \simeq 30\tau_c$. 
After this stage, both the electromagnetic energy conversion and temperature rising saturate asymptotically in the 2D simulation, with the system reaching a non-reconnecting relaxed state.
This corresponds to the end of plasmoid coalescence in the simulation, marked by the formation of a final magnetic island and the establishment of a new quasi-equilibrium  (see also the bottom-left panel of Fig. \ref{fig:FF_xy_plane}). 
Larger magnetic energy conversion is observed in model \texttt{ff3d} as in 3D, the energy injected into plasmoids can be further dissipated, whereas in 2D the plasmoids remain comparatively stable \citep{Sironi_Spitkovsky_2014, Sironi_Petropoulou_Giannios_2015}.
The conversion rate remains approximately constant up to $\tau_c \simeq 40$, after which it decreases with a corresponding counterpart in the temperature \citep[see also][]{Liu_2011}.
In model \texttt{ff3d} the dissipation of magnetic energy continues essentially for the whole duration of the simulation.


\begin{figure*}
    \centering
    \begin{minipage}{0.43\textwidth}
        \centering
        \includegraphics[width=\textwidth]{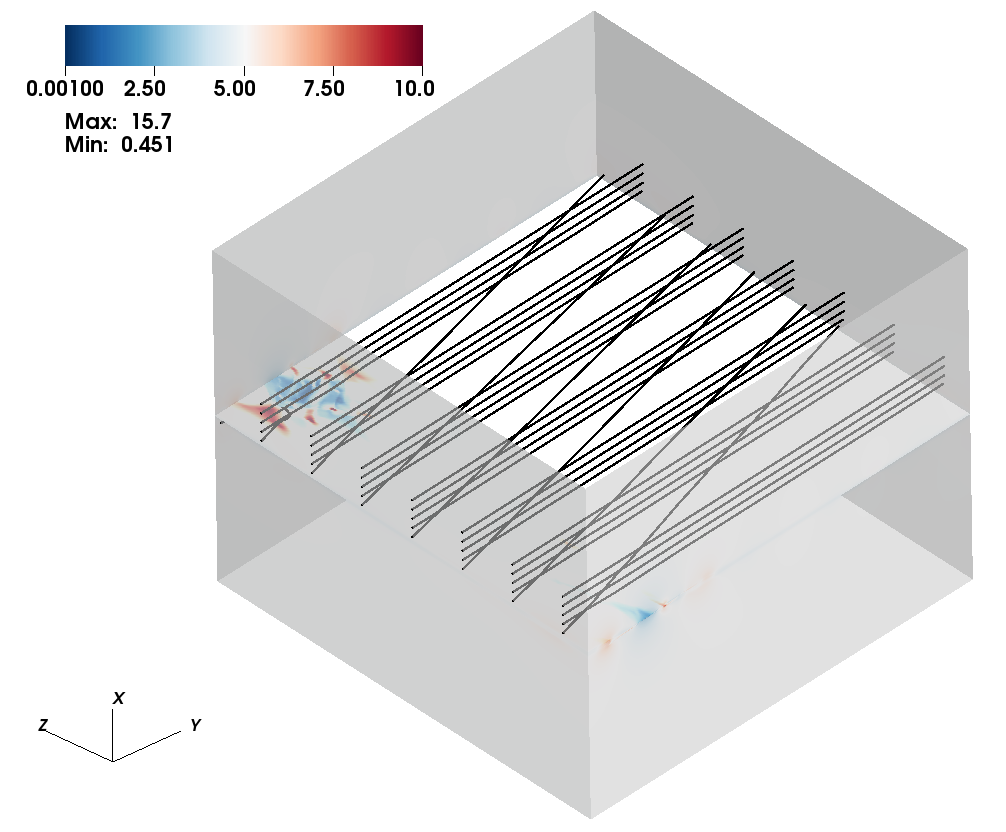}
        \subcaption{$\HALF B^2$ at $t = 10 \tau_c$}\label{fig:FF_B2_t10}
    \end{minipage} \hspace{1cm} \vspace{5mm}
    \begin{minipage}{0.40\textwidth}
        \centering
        \includegraphics[width=\textwidth]{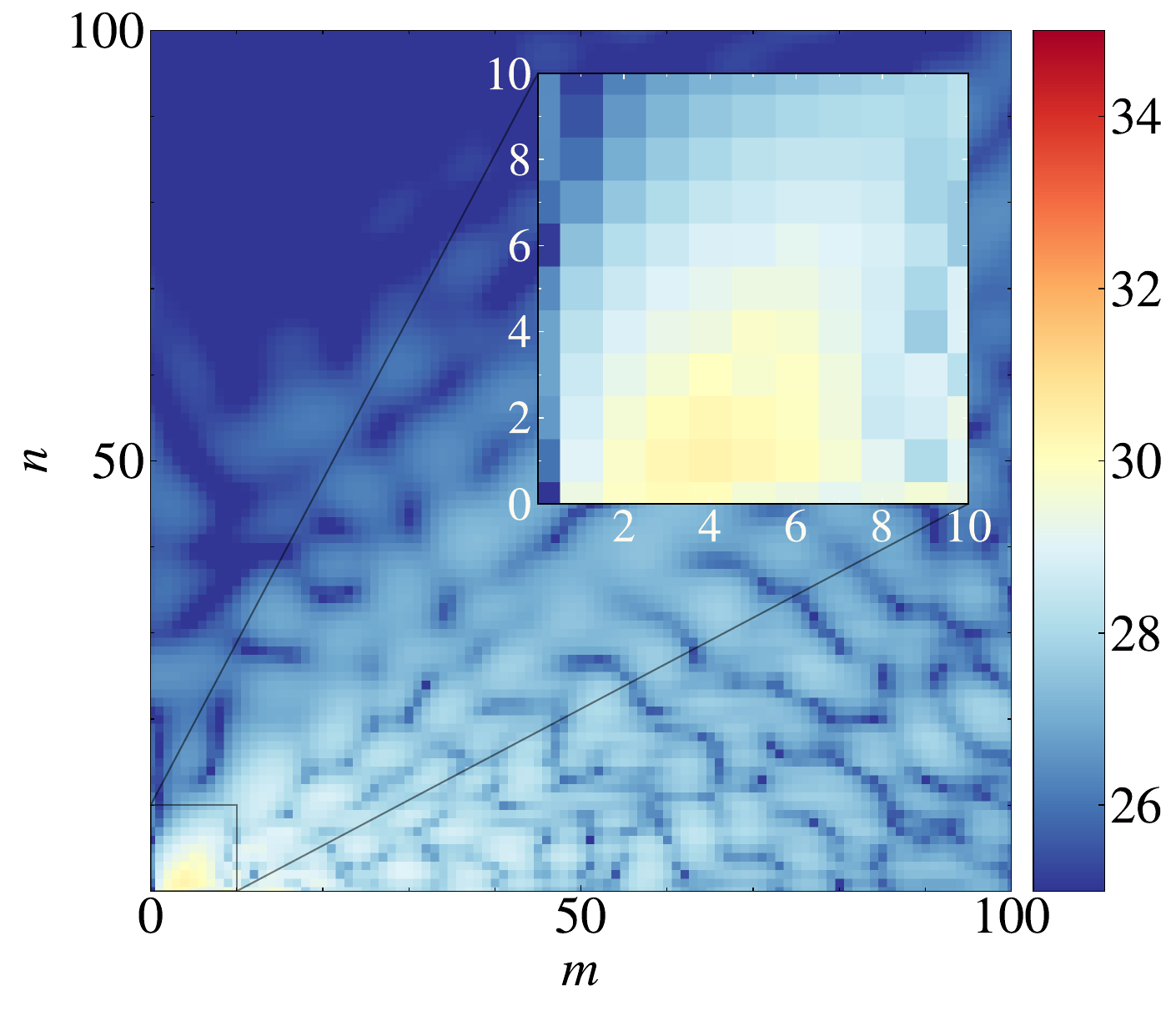}
        \subcaption{$\log(\mathcal{M}_{m,n}/\mathcal{M}_{0,0})$ at $t = 10 \tau_c$}\label{fig:FFT_t10}
    \end{minipage}

    \begin{minipage}{0.43\textwidth}
        \centering
        \includegraphics[width=\textwidth]{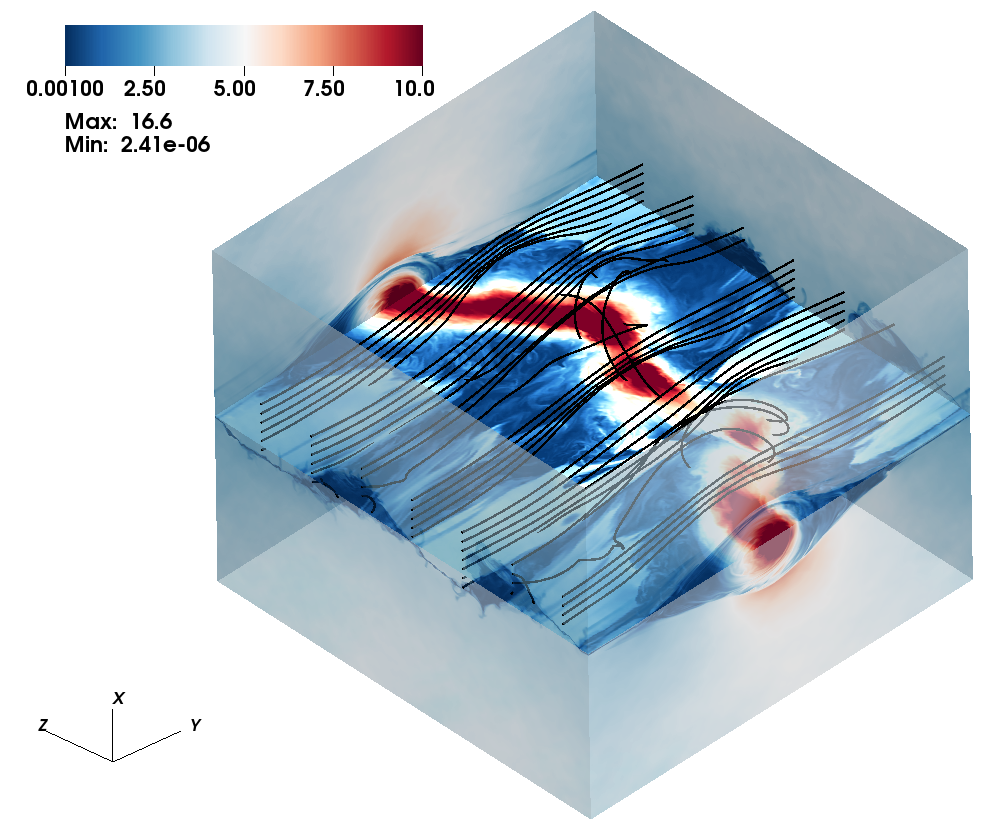}
        \subcaption{$\HALF B^2$ at $t = 20 \tau_c$}\label{fig:FF_B2_t20}
    \end{minipage} \hspace{1cm} \vspace{5mm}
    \begin{minipage}{0.40\textwidth}
        \centering
        \includegraphics[width=\textwidth]{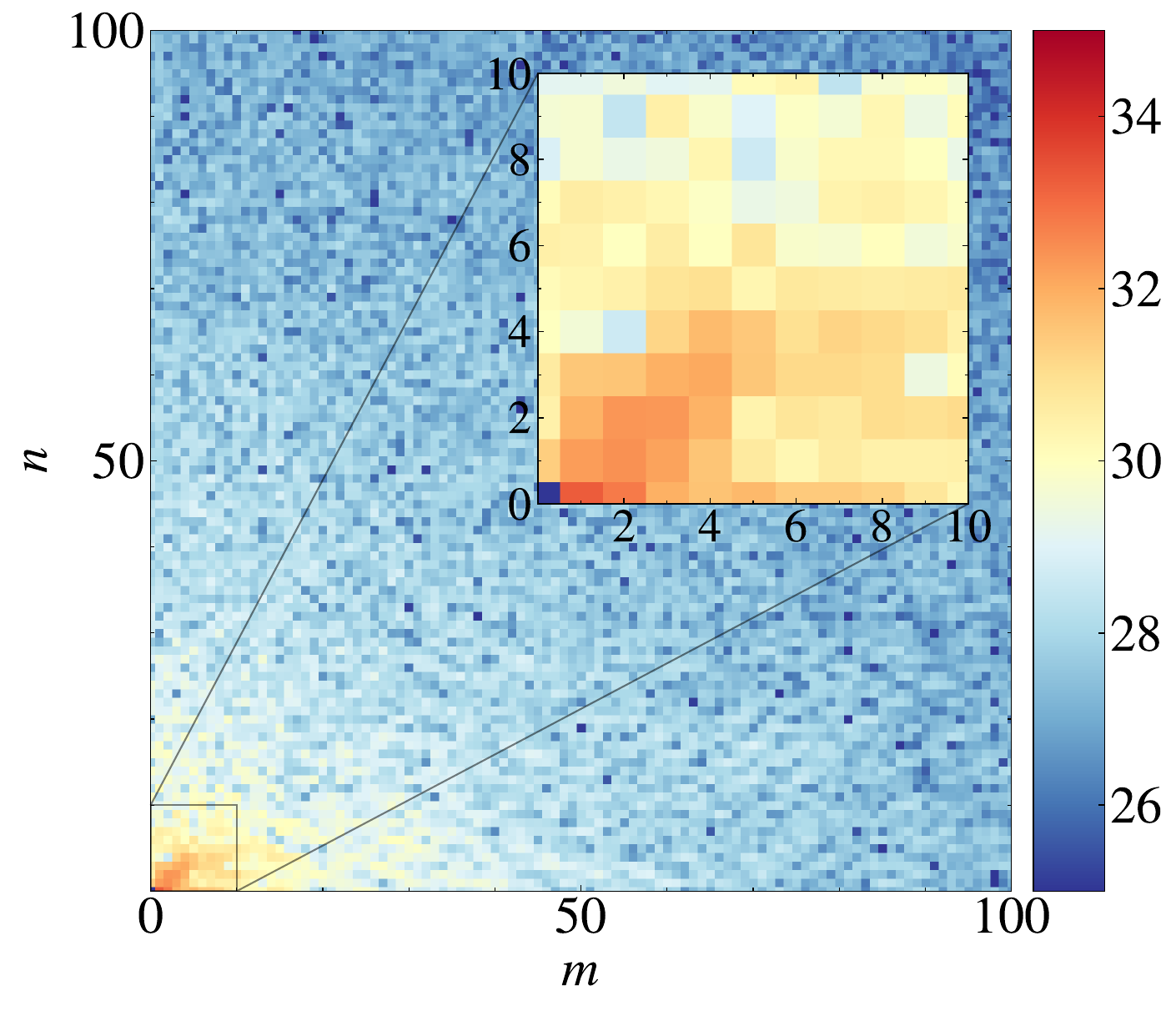}
        \subcaption{$\log(\mathcal{M}_{m,n}/\mathcal{M}_{0,0})$ at $t = 20 \tau_c$}\label{fig:FFT_t20}
    \end{minipage}

    \begin{minipage}{0.43\textwidth}
        \centering
        \includegraphics[width=\textwidth]{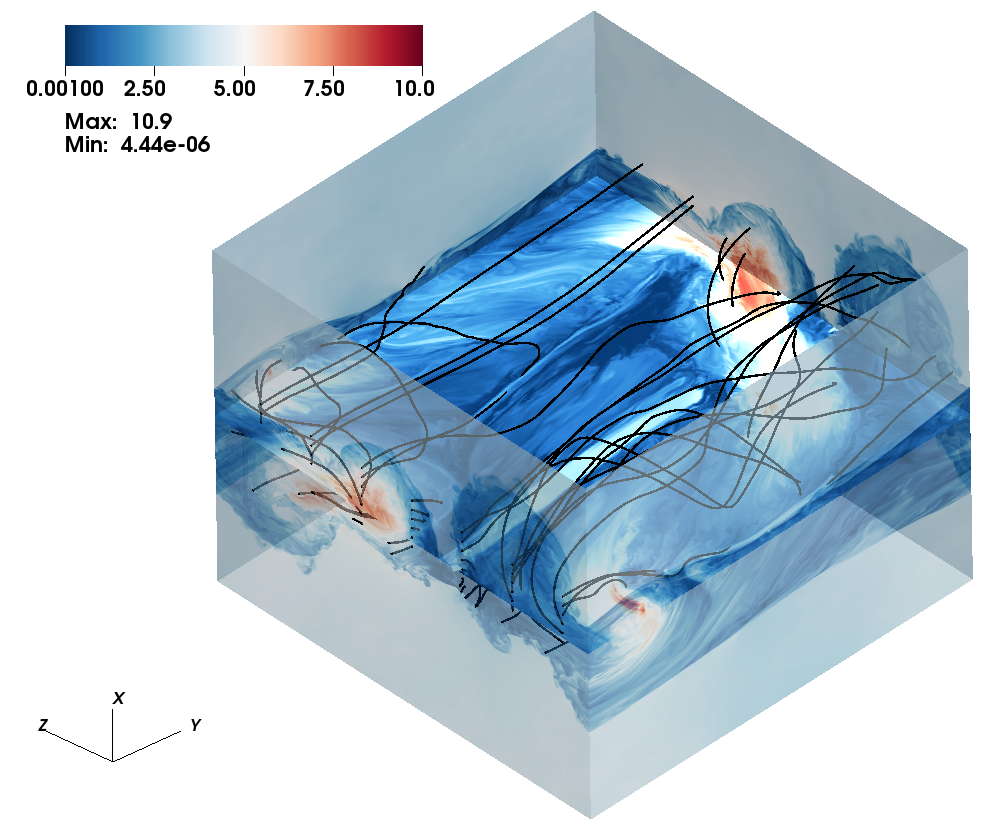}
        \subcaption{$\HALF B^2$ at $t = 35 \tau_c$}\label{fig:FF_B2_t35}
    \end{minipage} \hspace{1cm} \vspace{5mm}
    \begin{minipage}{0.40\textwidth}
        \centering
        \includegraphics[width=\textwidth]{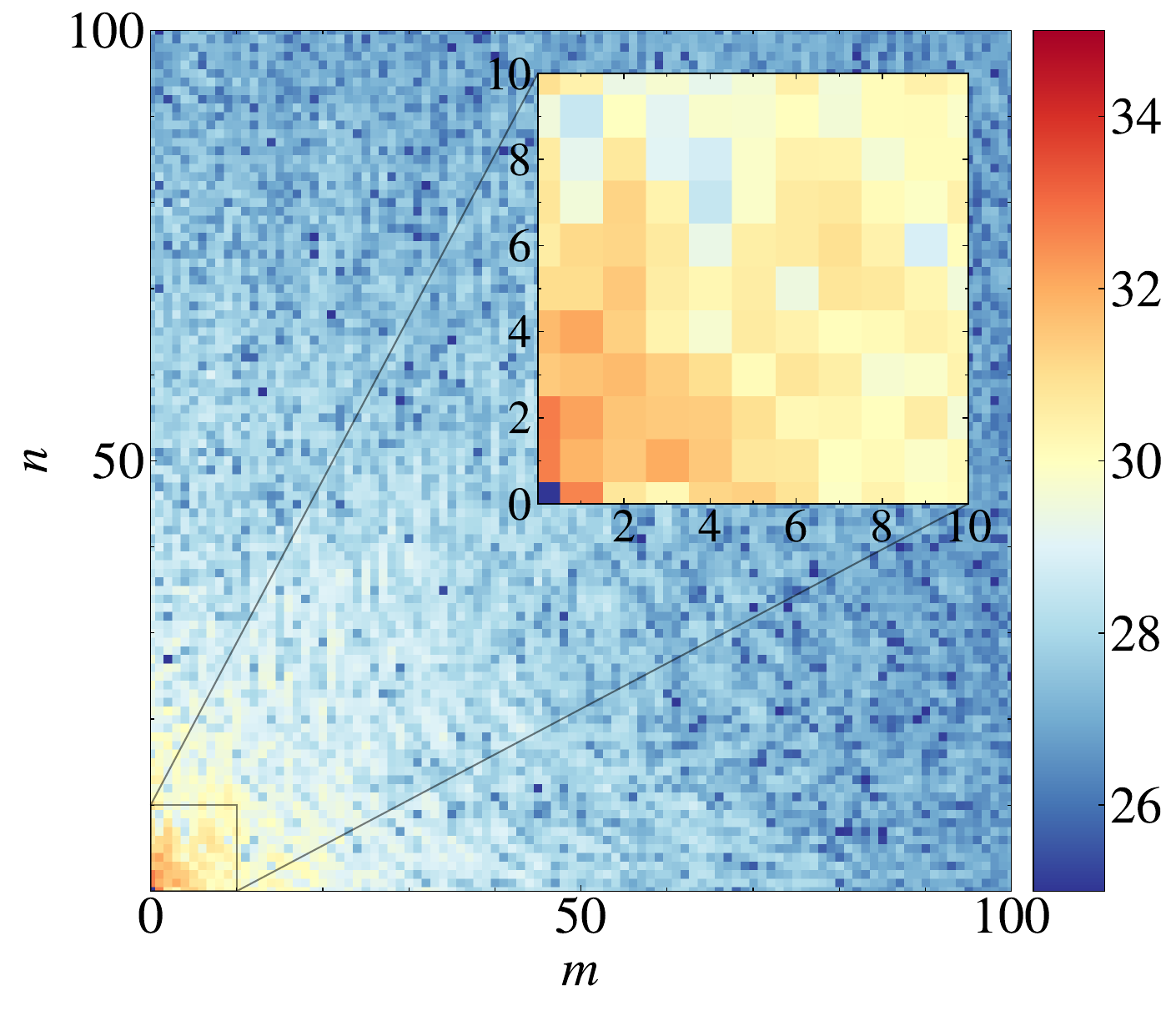}
        \subcaption{$\log(\mathcal{M}_{m,n}/\mathcal{M}_{0,0})$ at $t = 35 \tau_c$}\label{fig:FFT_t35}
    \end{minipage}

    \caption{\small Snapshots of the magnetic energy density ($\HALF B^2$, left column) with magnetic field lines overlaid in black at times $t = 10\tau_c$ (Fig. \ref{fig:FF_B2_t10}), $t = 20\tau_c$ (Fig. \ref{fig:FF_B2_t20}), and $t = 35\tau_c$ (Fig. \ref{fig:FF_B2_t35}). The corresponding Fourier transforms of the $x$-component of the magnetic field in the $x=0$ plane normalized by the $n=m=0$ mode ($\log(\mathcal{M}_{m,n}/\mathcal{M}_{0,0})$, right column) are shown in Figs. \ref{fig:FFT_t10}, \ref{fig:FFT_t20}, and \ref{fig:FFT_t35}. Wave-numbers are normalized to the box size ($L = 1$); for the considered grid resolution the numerical Nyquist limit corresponds to ($m,n \simeq 192$), above the maximum wave-numbers shown in the spectra. The zoom highlights the growth of the largest $10 \times 10$ modes.}
    \label{fig:FF_dyn}
\end{figure*}



\begin{figure*}
  \centering
  \includegraphics[width=0.95\textwidth]{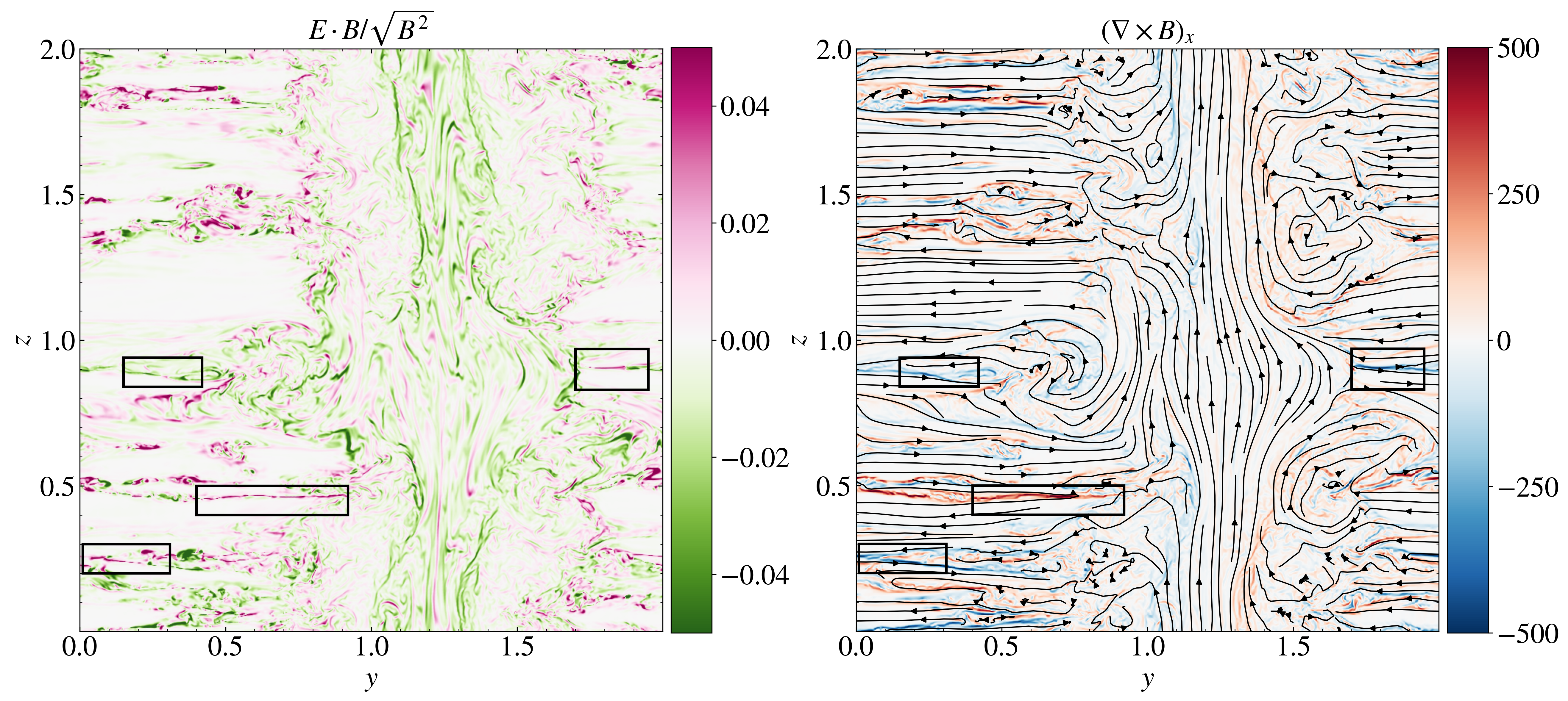}
  \caption{\small 2D slice cut on the current sheet plane $x = 0$ of the quantity $\vec{E} \cdot \vec{B}/B^2$ (left panel), and the $x$-component of $\nabla \times \vec{B}$ at time $t = 15\tau_{c}$ for data from model \texttt{ff3d}.
  Magnetic field streamlines are overlaid in black (right panel) for the proposed simulation, with black boxes emphasizing regions where secondary current sheets form.
}
  \label{fig:FF_current_sheets}
\end{figure*}


Having outlined the key differences and qualitative similarities between models with different dimensions, we now turn our attention to the 3D evolution of the current sheet.
The left column of Fig. \ref{fig:FF_dyn} shows snapshots of the magnetic energy density for model $\texttt{ff3d}$.
Snapshots are taken at times $t = 10\tau_c$ (Fig. \ref{fig:FF_B2_t10}), $t = 20\tau_c$ (Fig. \ref{fig:FF_B2_t20}), and $t = 35 \tau_c$ (Fig. \ref{fig:FF_B2_t35}) and show sliced cuts rendering of the current sheet plane and the boundary faces.
The right column of Fig. \ref{fig:FF_dyn} shows, instead, snapshots of the corresponding squared amplitude of the 2D Fourier transform of the of the $x$-component of the magnetic field in the $x=0$ plane  
\begin{equation}\label{eq:KH_fourier}
    \mathcal{M}_{m,n} = \left|\iint B_x e^{i(k_{y,m}y+k_{z,n}z)}dydz\right|^2 \, \,,
\end{equation}
normalized by the $n=m=0$ mode and also taken at times $t = 10\tau_c$ (Fig. \ref{fig:FFT_t10}), $t = 20\tau_c$ (Fig. \ref{fig:FFT_t20}), and $t = 35\tau_c$ (Fig. \ref{fig:FFT_t35}).

During the linear phase (Fig. \ref{fig:FF_B2_t10}) the various modes initially undergo an exponential growth characterized by an anisotropic spectrum, with long-wavelength modes, which possess the highest linear growth rates, preferentially excited along $y$ (Fig. \ref{fig:FFT_t10}).
This anisotropy can be attributed to the presence of an initial guide field along the $z$-direction, which introduces a magnetic tension on the current sheet's plane that inhibits and slows down the growth of modes along $z$.
Shortly after its onset, the tearing instability causes the initially uniform current layer to fragment into localized current-density blobs. 
As the system transitions into the non-linear regime, its evolution starts being governed by the dynamic coalescence of neighbouring magnetic islands. 
Both tearing and coalescence instabilities arise from the attractive force between parallel currents of opposite sign.
In close analogy with the 2D case, the 3D simulation shows that magnetic islands initially accelerate toward one another, driving flux compression and subsequent merging, such that the associated current-density structures grow progressively larger. 
Smaller magnetic islands are cannibalized, culminating in the formation of a dominant island. 
(see also Fig. \ref{fig:FF_xy_plane}).

While this configuration appears as the final quasi-stationary stage in our reference 2D simulation, the formation of a magnetized flux rope as a by-product of the coalescence instability in 3D follows a much richer evolution. 
In this phase, large-scale modes are preferentially excited, with $m = 1, 2$, and $n = 0,1$ dominating the spectrum (Fig. \ref{fig:FFT_t20}).
Cross-sectional views in the $z = 0$ and $z = 2L$ planes of Fig. \ref{fig:FF_B2_t20} show a plasmoid with the characteristic elongated, island-like shape associated with the tearing mode instability, with higher magnetic energy density within the O-point. 
The magnetic island extends along the $z$-direction, resembling a magnetized, current-carrying structure with quasi-elliptical cross-section, confined by a helical magnetic field whose tension is balanced by the pressure at the O-point. 
Once the tearing-generated flux rope becomes sufficiently large, aided by the stabilizing effect of $B_z$,
it becomes unstable to current-driven modes, leading to the twisting and kink-like distortion observed across the current sheet plane (Fig. \ref{fig:FF_B2_t20}). 

Both the coalescence instability and the kinking of the flux rope are parasitic instabilities of tearing and promote the emergence of structures in the $z$-direction, favouring the onset of fully 3D turbulence in the computational box. 
Their subsequent non-linear evolution reduce the coherence length of magnetic and plasma structures along the $z$-direction, opening new channels for field dissipation.
In fact, by $t = 35\tau_c$ (Fig. \ref{fig:FF_B2_t35}), the flux rope is fully disrupted, with emerging modes and structures propagating in the $z$-direction. 
Comparing Figs. \ref{fig:FFT_t20}–\ref{fig:FFT_t35}, the centroid of the mode distribution has gradually shifted in the $z$-direction, indicating the slower but persistent growth of perturbations along this axis. 
This disruption promotes the onset of self-generated turbulence and the thickening of the reconnection layer. 
This result is consistent with first-principle kinetic models which likewise exhibit a chaotic, 3D turbulent evolution \citep[see, e.g.,][and references therein]{Werner_Uzdensky_2021}.

As a last consideration, during the non-linear phase of the tearing dynamics, the onset of parasitic instabilities foster additional secondary reconnection events.
Fig. \ref{fig:FF_current_sheets} presents a 2D slice of the current-sheet plane at $t = 15\tau_{c}$, showing two complementary diagnostics of non-ideal activity within the reconnecting layer. 
The left panel displays the quantity $\mathbf{E}\cdot\mathbf{B}/\sqrt{B^{2}}$, clipped to the interval $[-0.05,0.05]$ to enhance the visibility of regions where ideal MHD is violated. 
The right panel shows the $x$-component of the current-density proxy $(\nabla\times\mathbf{B})_{x}$, spanning the range $[-500,500]$.
Overlaid black streamlines trace the in-plane magnetic field and reveal the local topology. 
In both panels, black rectangles mark representative patches where the field reverses polarity and where the diagnostics indicate enhanced activity.

The two diagnostics jointly map the intricate structure of the evolving current sheet. 
The regions highlighted by the rectangles correspond to thin, elongated current sheets that pierce the cross-section and are associated with departures from the ideal RMHD regime. 
High values of $\mathbf{E}\cdot\mathbf{B}/\sqrt{B^2}$ correspond to sites where field lines annihilate, while peaks in $(\nabla\times\mathbf{B})_{x}$ trace intense, filamentary currents produced by the non-linear evolution of the tearing. 
In the interval $y\in[1.0,1.5]$, the current distribution outlines the core of the flux rope, whose magnetic field lines run predominantly along the $z$-direction. 
Together, these features demonstrate that the disruption of the primary current layer in 3D seeds multiple secondary reconnection events: local polarity reversals, strong currents, and enhanced non-ideal electric fields all indicate active tearing and fragmentation of newly formed current sheets.

\subsection{Pressure-balanced configuration}
\label{sec:PB}
%

\begin{figure}
  \centering
  \includegraphics[width=0.45\textwidth]{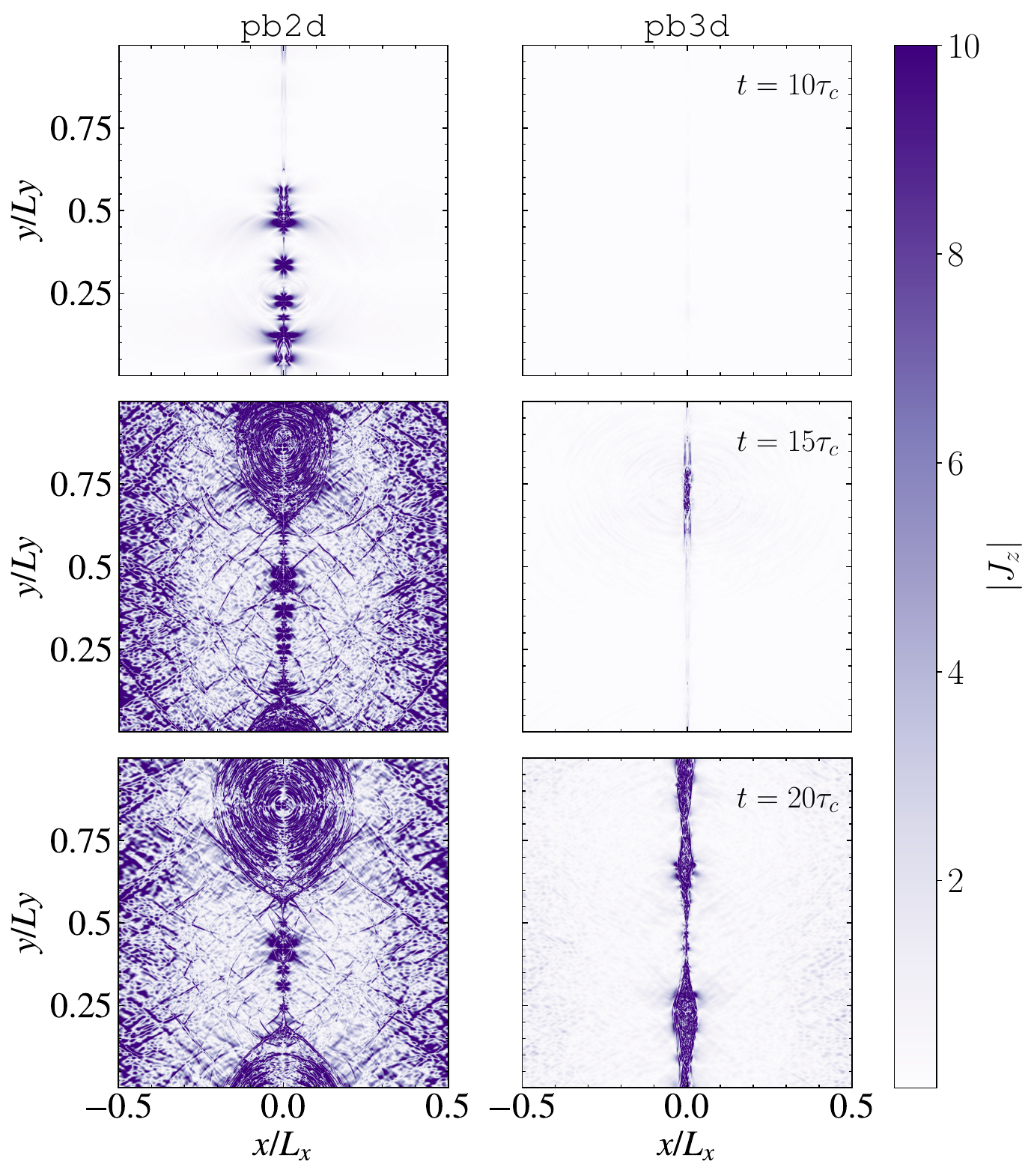}
  \caption{\small Snapshot of the module of the $z$-component of $|\nabla \times \vec{B}|$ used as a proxy for $|J_z|$ in the plane $z = 0$ for model \texttt{pb2d} (left column), and module \texttt{pb3d} (right column) at three different times: $t = 10\tau_c$ (top row), $t = 15\tau_c$ (middle row), $t = 20\tau_c$ (bottom row).   
}
  \label{fig:PB_xy_plane}
\end{figure}

We now turn to the pressure-balanced configuration, comparing models \texttt{pb2d} and \texttt{pb3d}.
%
%
\begin{figure}
  \centering
  \includegraphics[width=0.45\textwidth]{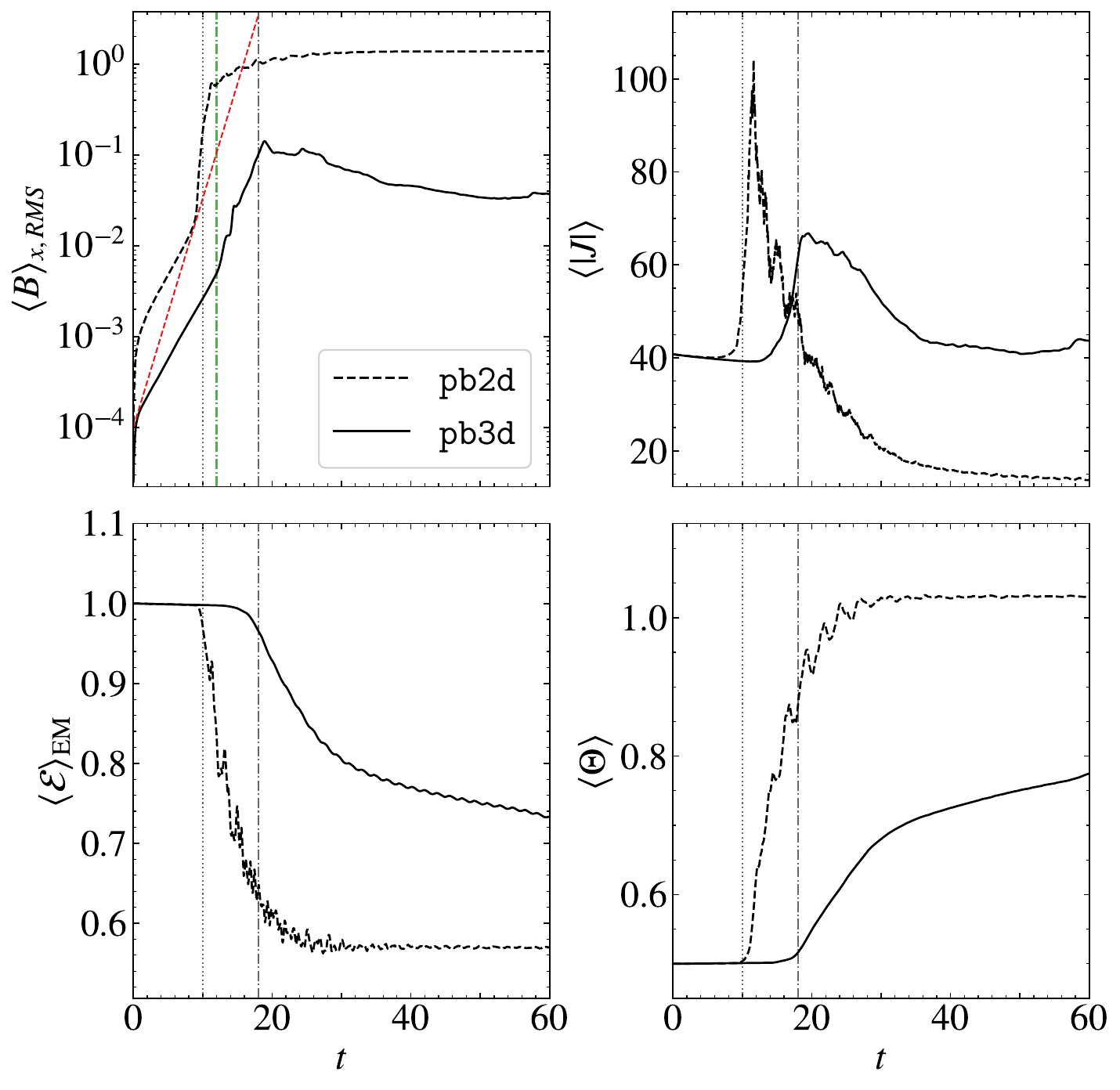}
  \caption{\small Temporal evolution of the root-mean-square of the magnetic field component $B_{x}$ (upper left panel), the module of $|\nabla \times \vec{B}|$ as proxy for the non-relativistic electric current $|J|$ (upper right), electromagnetic energy normalized to its initial value ${\cal E}_{\text{EM}} =(E^2 + B^2)/(B_0^2)$ (lower left), and the average relativistic temperature ${\Theta} = p/\rho$ (lower right) for simulations initialized with the pressure-balanced equilibrium. All quantities integrated over the uniform portion of the computational volume are indicated with the $\av{}$ symbol.
  Time is given in units of the light-crossing time of half the sheet length, $\tau_c = L/c$.
  The black dotted and dash-dotted vertical lines mark, respectively, the transition to the non-linear regime. In the upper-left panel, the green dash-dotted line identifies the time at which the linear phase terminates in the corresponding force-free reference run, and the red dashed line marks the slope of the linear growth rate of ideal tearing in models \texttt{ff2d} and \texttt{ff3d}.
}
  \label{fig:PB_time_evolution}
\end{figure}
%
%
The same set of diagnostics used in the force-free analysis is shown in Figs. \ref{fig:PB_time_evolution} and \ref{fig:PB_xy_plane}.
%
%
%
%
As shown in Fig. \ref{fig:PB_time_evolution}, models \texttt{pb2d} and \texttt{pb3d} exhibit markedly different time evolutions, both with respect to each other and relative to the force-free models \texttt{ff2d} and \texttt{ff3d}.
In planar symmetry, the linear stage exhibits the expected behaviour of reconnection mediated by ideal tearing: an initial linear growth of tearing, followed by a linear phase of plasmoid coalescence (see upper-left panel of Fig. \ref{fig:PB_time_evolution}). 
The associated peak in $\av{|J|}$ (upper-right panel of Fig. \ref{fig:PB_time_evolution}) mirrors this behaviour and signals the onset of a major reconnection event in 2D.
In model \texttt{pb2d}, the quantity $\av{B}_{x,\mathrm{RMS}}$ reaches the end of its exponential growth around $t \simeq 10-12\tau_{c}$, as indicated by the black dotted vertical line in the upper-left panel of Fig. \ref{fig:PB_time_evolution}.
The dashed red line overlaid on the same panel indicates the growth rate of the ideal tearing instability during its linear phase (i.e., $t \simeq 2-6\tau_{c}$) in models \texttt{ff2d} and \texttt{ff3d}.
We note that the slope of the linear phase of tearing is remarkably different between the two equilibria, resulting in model \texttt{pb2d} exiting the linear phase slightly later than the corresponding force-free case, which does so at $t \simeq 8\tau_{c}$.
On the other hand, model \texttt{pb3d} initially follows the exponential growth of tearing instability up to $t \simeq 12\tau_{c}$ (green dash-dotted vertical line in the upper-left panel of Fig. \ref{fig:PB_time_evolution}) with the same slope of the 2D case.
It is worth noting that by the same time model \texttt{ff3d} has also reached the end of the exponential growth of the field.
After $t \simeq 12\tau_{c}$, the two trends become notably different, as the growth of the plasmoid coalescence is quenched in model \texttt{pb3d}, suggesting that the 3D simulation does not develop the instability as in planar symmetry.
The subsequent dynamics are sensibly different. 
In fact, beyond this point the growth of the reconnected magnetic field in model \texttt{pb3d} is manifestly hindered and reaches only a shallow maximum around $t \simeq 18\tau_{c}$ (black dash-dotted vertical line in Fig. \ref{fig:PB_time_evolution}), without displaying the characteristic rise-and-saturation profile which follows plasmoid coalescence.
Consistently, the current spike observed in model \texttt{pb2d} (upper-right panel) is visibly quenched in model \texttt{pb3d}.
The behaviour of the energy diagnostics shown in the bottom panels of Fig. \ref{fig:PB_time_evolution} provides additional evidence that reconnection and plasmoid coalescence are suppressed in model \texttt{pb3d}.
While the 2D model undergoes a conversion of magnetic energy into thermal energy analogous to the respective force-free case, the 3D run shows only a modest decrease in $\av{\mathcal{E}}_{\mathrm{EM}}$ and a correspondingly weaker increase in temperature. 

Spatial information depicting the overall dynamics described above is provided in Fig. \ref{fig:PB_xy_plane}, which displays snapshots of the magnitude of the $z$-component of $|\nabla \times \vec{B}|$ as proxy for $|J_z|$ in the $z = 0$ plane.
As in the force-free case, the left and right columns correspond to the 2D and 3D models, respectively, and the three rows illustrate successive stages in the early evolution.
In this case, snapshots are taken at times $t = 10\tau_c$ (top), $t = 15\tau_c$ (middle), and $t = 20\tau_c$ (bottom). 
At $t=10\tau_c$, the 2D simulation shows the emergence of the reconnecting current layer characterized by the expulsion of plasmoids as a by-product of ongoing reconnection.
By $t=15\tau_c$, the linear phase of the plasmoid coalescence has terminated in model \texttt{pb2d}, and the system has entered a non-linear regime of current sheet fragmentation.
At the corresponding time, in the 3D simulation, plasmoids are forming in the upper region of the current layer (mid-right panel of Fig. \ref{fig:PB_xy_plane}).
At $t = 20\tau_{c}$, the 2D model displays the familiar fully developed non-linear state dominated by a forming final magnetic island resulting from repeated tearing and coalescence. 
In 3D, instead, the sheet does indeed start reconnecting, but the characteristic plasmoid coalescence instability driven by tearing is damped: the current layer thickens and becomes more irregular, with elongated, irregular plasmoids forming along the sheet. 
Together, the trends shown in Figs. \ref{fig:PB_time_evolution} and \ref{fig:PB_xy_plane} suggest that model \texttt{pb3d} experiences tearing, but the plasmoid-mediated reconnection is quenched.


\begin{figure*}
    \centering
    \begin{minipage}{0.43\textwidth}
        \centering
        \includegraphics[width=\textwidth]{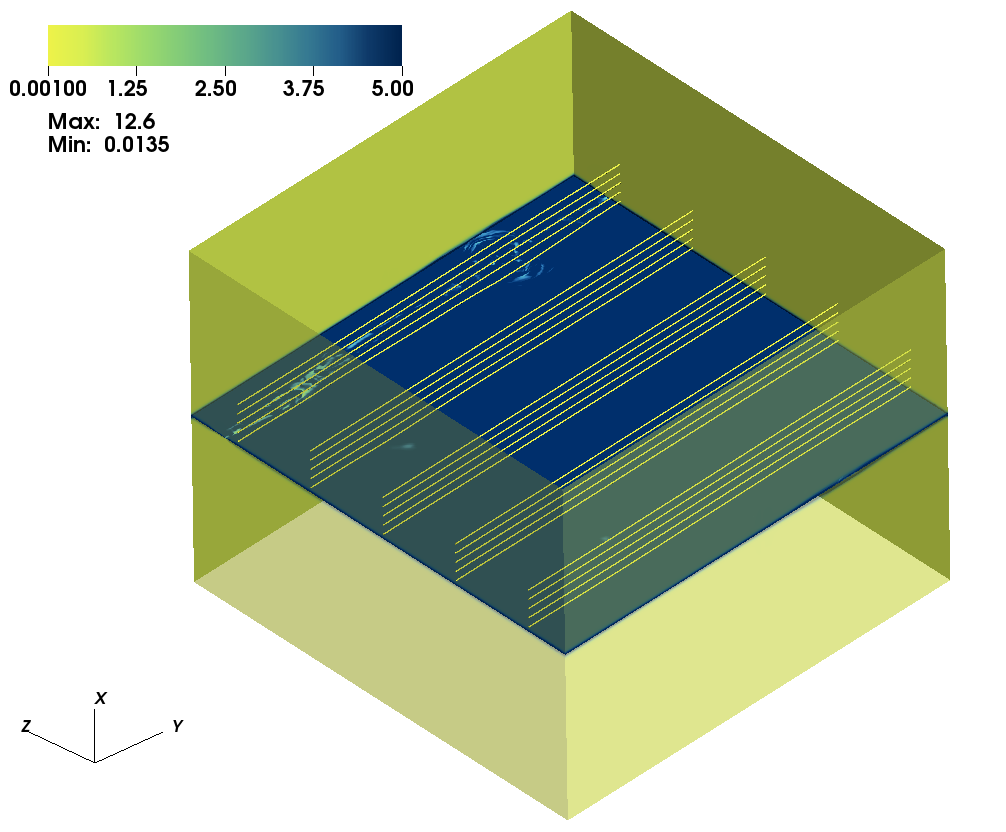}
        \subcaption{$p$ at $t = 15 \tau_c$}\label{fig:PB_prs_15}
    \end{minipage} \hspace{1cm} \vspace{5mm}
    \begin{minipage}{0.40\textwidth}
        \centering
        \includegraphics[width=\textwidth]{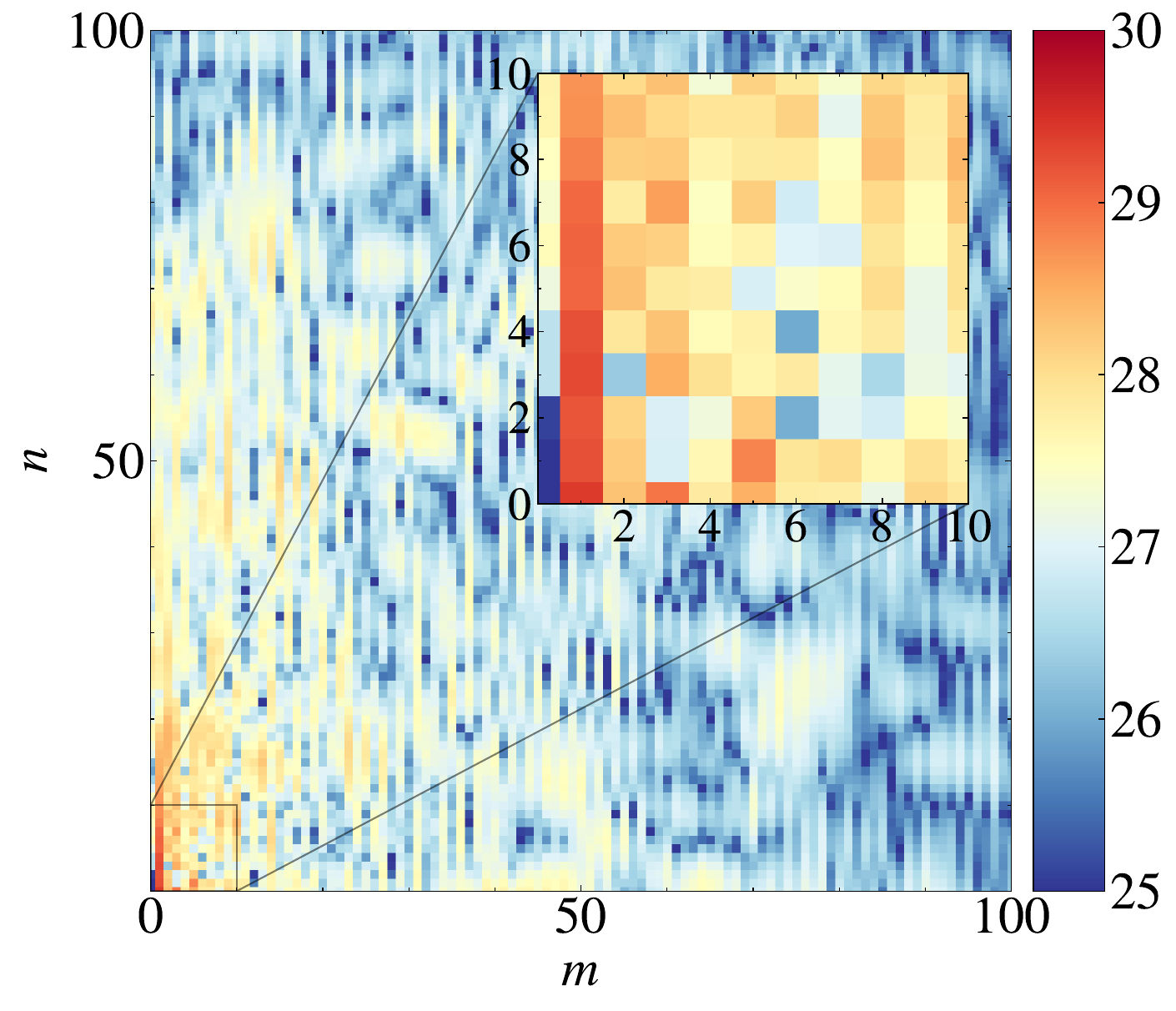}
        \subcaption{$\log(\mathcal{M}_{m,n}/\mathcal{M}_{0,0})$ at $t = 15 \tau_c$}\label{fig:PB_FFT_30}
    \end{minipage}

    \begin{minipage}{0.43\textwidth}
        \centering
        \includegraphics[width=\textwidth]{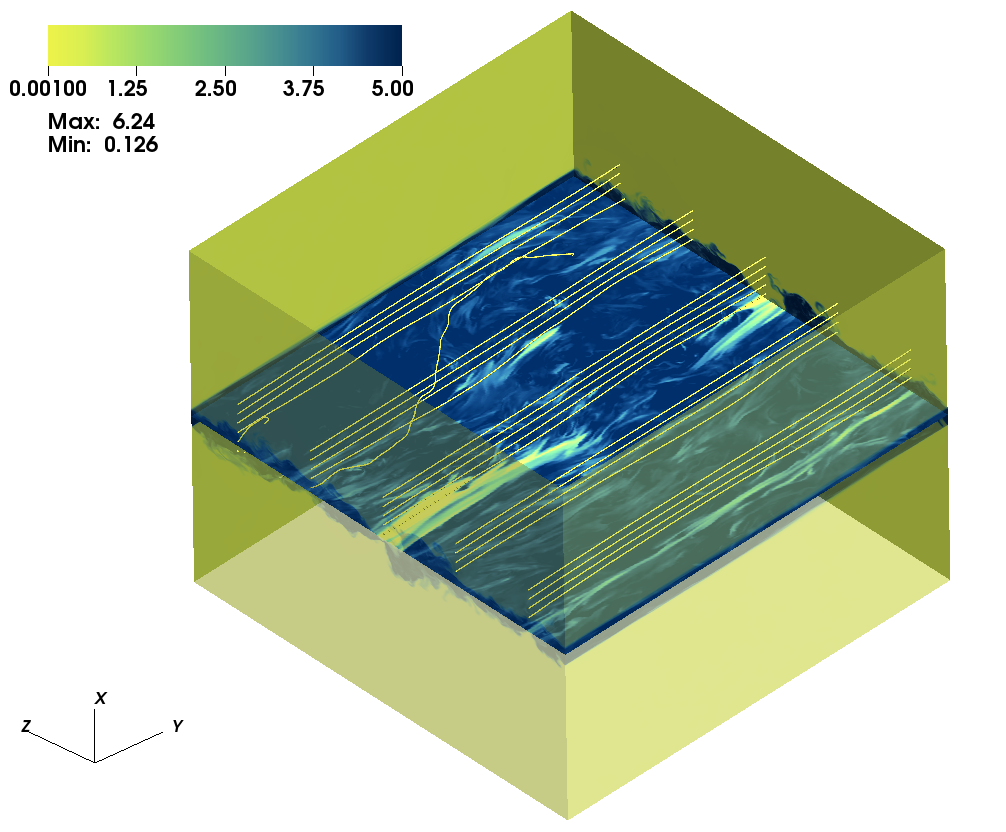}
        \subcaption{$p$ at $t = 30 \tau_c$}\label{fig:PB_prs_t30}
    \end{minipage} \hspace{1cm} \vspace{5mm}
    \begin{minipage}{0.40\textwidth}
        \centering
        \includegraphics[width=\textwidth]{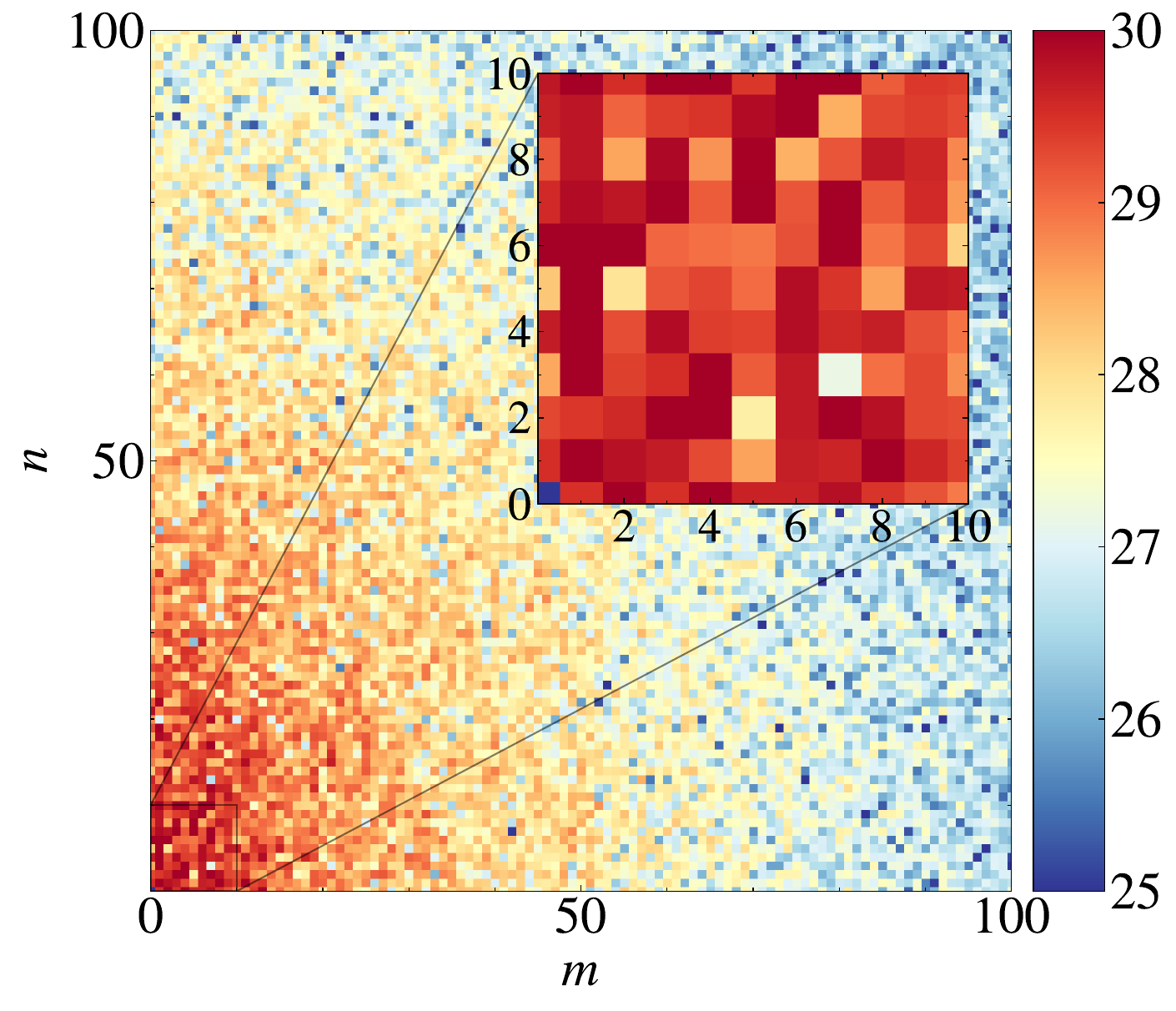}
        \subcaption{$\log(\mathcal{M}_{m,n}/\mathcal{M}_{0,0})$ at $t = 30 \tau_c$}\label{fig:PB_FFT_60}
    \end{minipage}

    \begin{minipage}{0.43\textwidth}
        \centering
        \includegraphics[width=\textwidth]{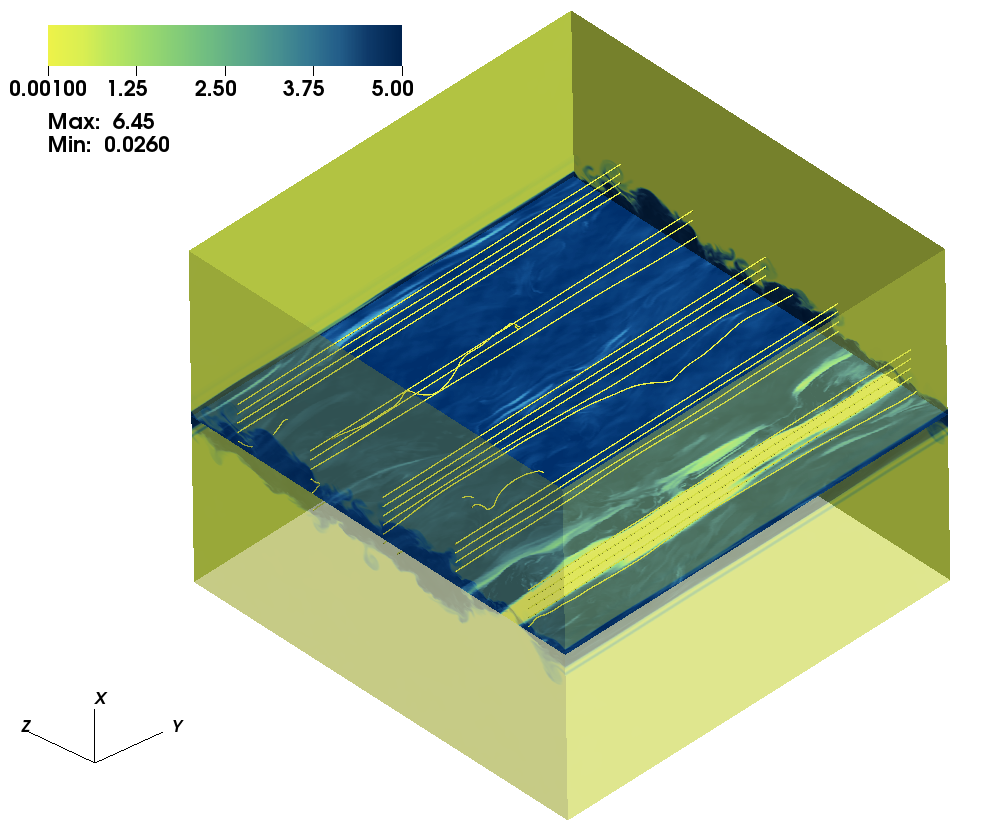}
        \subcaption{$p$ at $t = 60\tau_c$}\label{fig:PB_prs_t60}
    \end{minipage} \hspace{1cm} \vspace{5mm}
    \begin{minipage}{0.40\textwidth}
        \centering
        \includegraphics[width=\textwidth]{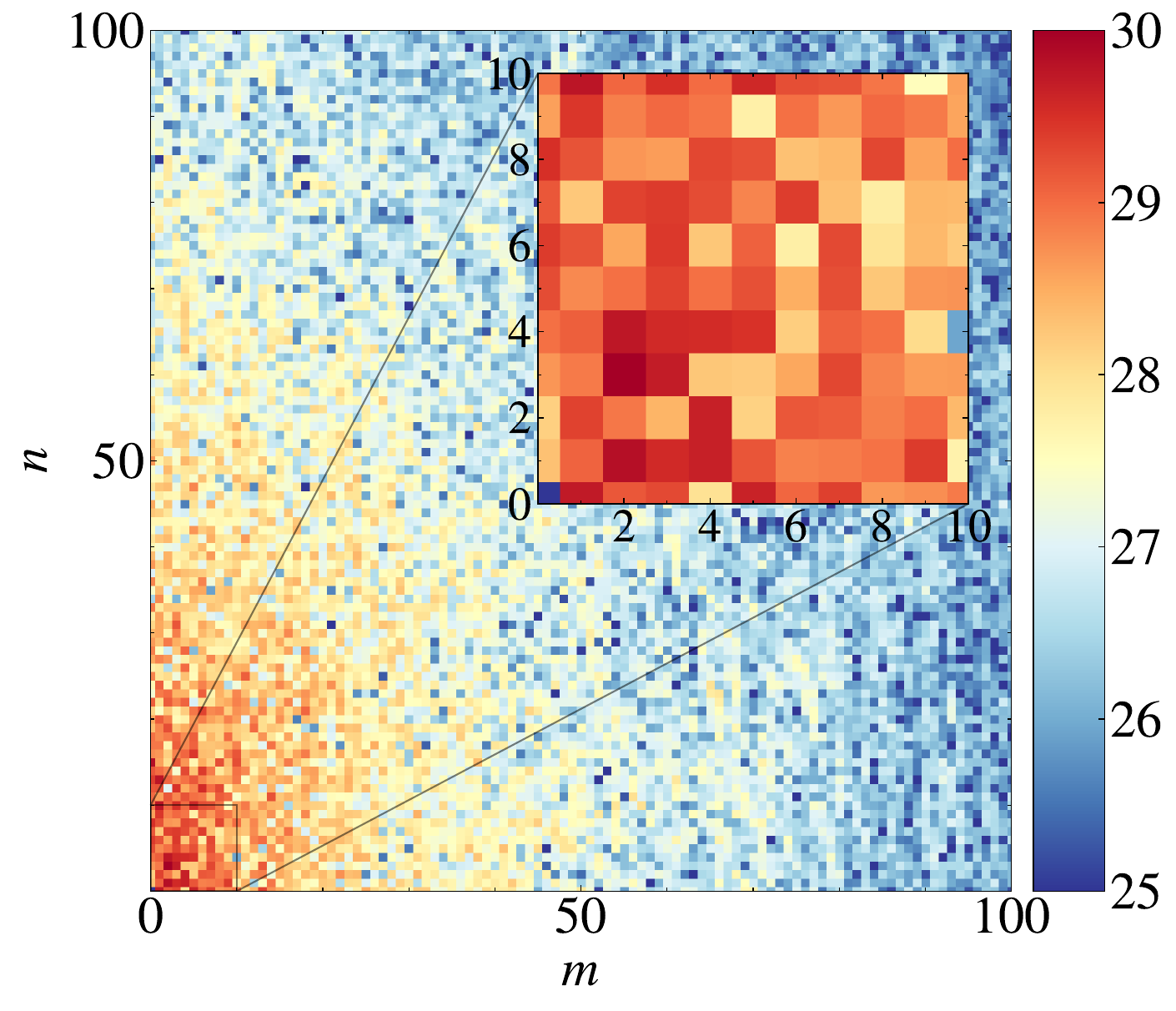}
        \subcaption{$\log(\mathcal{M}_{m,n}/\mathcal{M}_{0,0})$ at $t = 60 \tau_c$}\label{fig:PB_FFT_120}
    \end{minipage}

    \caption{\small Snapshots of the thermal pressure ($p$, left column) with magnetic field lines overlaid in yellow at times $t = 15\tau_c$ (Fig. \ref{fig:PB_prs_15}), $t = 30\tau_c$ (Fig. \ref{fig:PB_prs_t30}), and $t = 60\tau_c$ (Fig. \ref{fig:PB_prs_t60}). Corresponding Fourier transforms of the $x$-component of the magnetic field ($\log(\mathcal{M}_{m,n}/\mathcal{M}_{0,0})$, right column) are shown in Figs. \ref{fig:PB_FFT_30}, \ref{fig:PB_FFT_60}, and \ref{fig:PB_FFT_120}. Wave-numbers are normalized to the box size ($L = 1$); for the considered grid resolution the numerical Nyquist limit corresponds to ($m,n \simeq 192$), above the maximum wave-numbers shown in the spectra. The zoom highlights the growth of the first $10 \times 10$ modes.}
    \label{fig:PB_dyn}
\end{figure*}


To understand the reason behind the quenching of the reconnection dynamics in model \texttt{pb3d}, Fig. \ref{fig:PB_dyn} illustrates the dynamical evolution of the pressure-balanced current sheet at three representative stages of the simulation, together with the corresponding bidimensional Fourier spectra of the reconnected magnetic field component as in Eq. (\ref{eq:KH_fourier}). 
In this case, the left column of Fig. \ref{fig:PB_dyn} shows snapshots of the thermal pressure for model $\texttt{pb3d}$.
Snapshots are taken at times $t = 15\tau_c$ (Fig. \ref{fig:PB_prs_15}), $t = 30\tau_c$ (Fig. \ref{fig:PB_prs_t30}), and $t = 60 \tau_c$ (Fig. \ref{fig:PB_prs_t60}) and show sliced cuts rendering of the current sheet plane and the boundary faces.

At $t = 15\tau_{c}$ (Figs. \ref{fig:PB_prs_15} and \ref{fig:PB_FFT_30}), the thermal-pressure is higher on the current sheet than in the upstream plasma due to the initial equilibrium configuration.
Some signs of reconnection along the current sheet are visible (see also the $z = 0$ cut in the mid-right panel of Fig. \ref{fig:PB_xy_plane} at the corresponding time). 
The plasma $\beta$ on the current sheet is very high ($\beta_{\rm min} \simeq \beta_{\rm max} \simeq 10^2$ at $t = 0$, and $\beta_{\rm min} \simeq 10^{-1}$ while $\beta_{\rm max} \simeq 10^7$ at $t = 15 \tau_c$).
High $\beta$ values on the current sheet favour the onset of pressure-driven modes: as $\beta$ increases, the destabilizing effect of local pressure gradients becomes stronger relative to the magnetic tension that stabilizes the system, making it easier for perturbations to grow in regions of unfavourable curvature \citep{Biskamp_1986}. 
In our configuration, the magnetic tension is weak due to the absence of a guide field component ($B_z = 0$) in the initial setup. 
Consequently, pressure-driven instabilities can compete with or alter other dynamics, such as reconnection, by rapidly acting on the formed plasmoids, which are destroyed before they can coalesce.
This dynamics is suggested also by the Fourier spectrum at this stage, which is dramatically different from that of model \texttt{ff3d} at similar times (compare Fig. \ref{fig:PB_FFT_30} with Fig. \ref{fig:FFT_t10}).
Fig. \ref{fig:PB_FFT_30} shows the excitation of the $m = 1$ mode together with a broad range of low- to intermediate-$z$ wavelengths, as evidenced by the dominance of modes $\mathcal{M}_{1,n}$ with $n \lesssim 30$, which are a signature of emerging pressure-driven modes \citep{Strauss_1981, Biskamp_MagRec}. 
The suppression of the second linear phase, associated with the linear growth of the plasmoid coalescence instability in the upper panel of Fig. \ref{fig:PB_time_evolution}, along with the irregular morphology of plasmoids visible in the lower-right panel of Fig. \ref{fig:PB_xy_plane}, corroborates this dynamics.
The absence of a guide field in the $z$-direction in the 3D setup is crucial for establishing this behaviour, along with the role of dimensionality. 
In fact, in the 2D pressure-balanced configuration, this effect cannot occur, since the lack of $z$-dependence precludes the growth of modes with $n > 0$, thus leading the planar symmetry to mimic the stabilizing effect of a guide field.

By $t = 30\tau_{c}$, (Figs.\ref{fig:PB_prs_t30} and \ref{fig:PB_FFT_60}), model \texttt{pb3d} has entered a transitional regime where the tearing forms plasmoids but fails to produce a coherent, dominant reconnection structures such as the flux tube observed in model \texttt{ff3d}. 
The pressure distribution exhibits elongated, wavy-like distortions along the current layer. 
In Fourier space, power has now spread across a broader region of the $(m,n)$ plane: a wide range of small-to-intermediate scale modes grow above the numerical noise. 
The spectrum lacks a single dominant mode.
By the end of the simulation ($t = 60\tau_{c}$, Figs. \ref{fig:PB_prs_t60}, and \ref{fig:PB_FFT_120}), the system has transitioned into a non-linear, turbulent-like regime causing deformation and thickening of the current layer. 
The layer is populated by small-scale structures and weak filamentary features in the $z$-direction. 
Magnetic field lines show stochastic wandering, suggesting the development of 3D shear-driven distortions. 
The Fourier spectrum at this stage reflects this behaviour: power is widely distributed, with no pronounced ridge or peak.

\subsection{Non-ideal regions}
\label{sec:NonIdeal}

%

\begin{figure}
  \centering
  \includegraphics[width=0.45\textwidth]{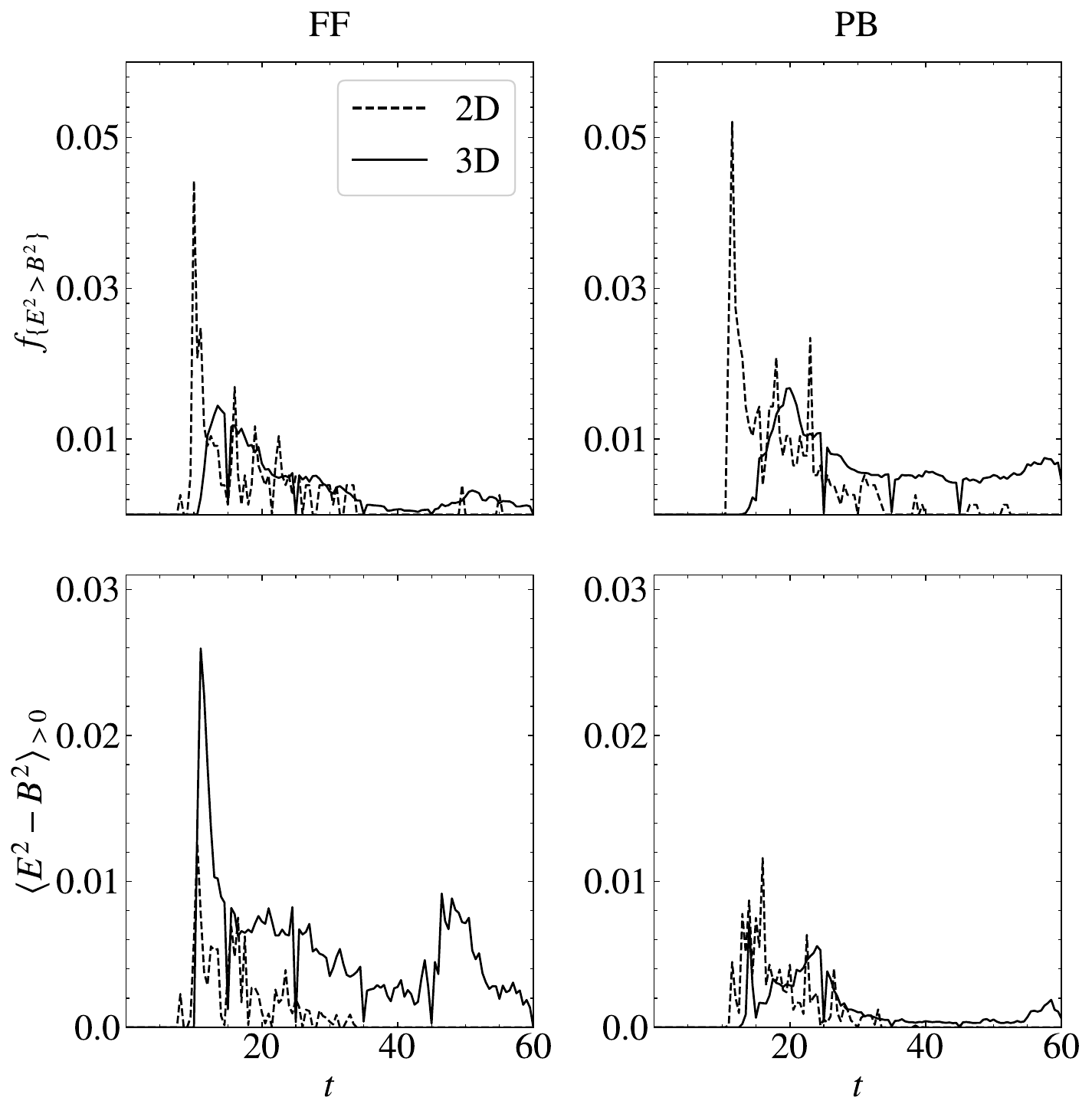}
  \caption{\small On the top row is shown the temporal evolution of the overall fraction of grid 'pencils' along the x-direction having at least one grid cell with electric dominance (i.e. $E^2>B^2$).
  On the bottom row instead there is the temporal evolution of the volume(surface)-integrated mean value of the Lorentz invariant $ \langle E^2 - B^2 \rangle $, averaged over regions where $ E^2 - B^2$ is positive.
  The left column confront 2D (dashed lines) vs 3D (solid lines) force-free cases, while the right column the pressure balanced cases.
}
  \label{fig:E2B2}
\end{figure}



\begin{figure*}
    \centering
    \begin{minipage}{0.44\textwidth}
        \centering
        \includegraphics[width=\textwidth]{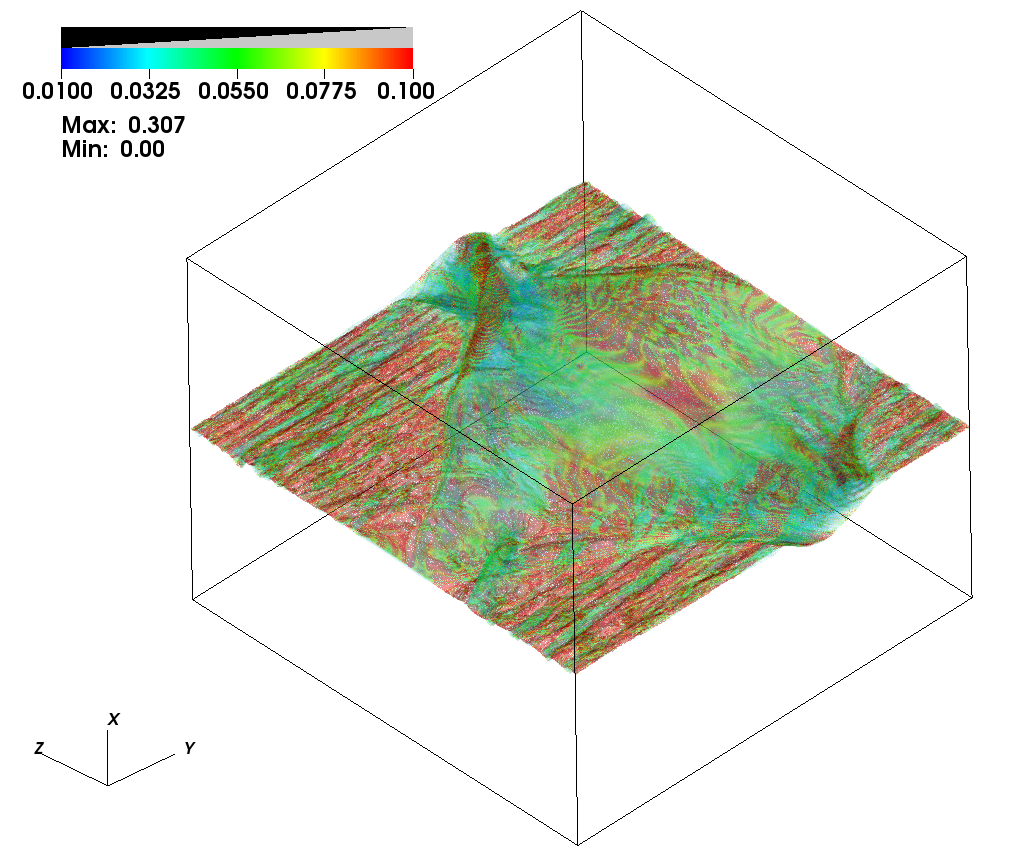}
        \subcaption{$E^{\rm res}_{\rm FF}$ at $t = 11.5 \tau_c$}\label{fig:nonid_FF_23}
    \end{minipage} \hspace{1cm} \vspace{5mm}
    \begin{minipage}{0.44\textwidth}
        \centering
        \includegraphics[width=\textwidth]{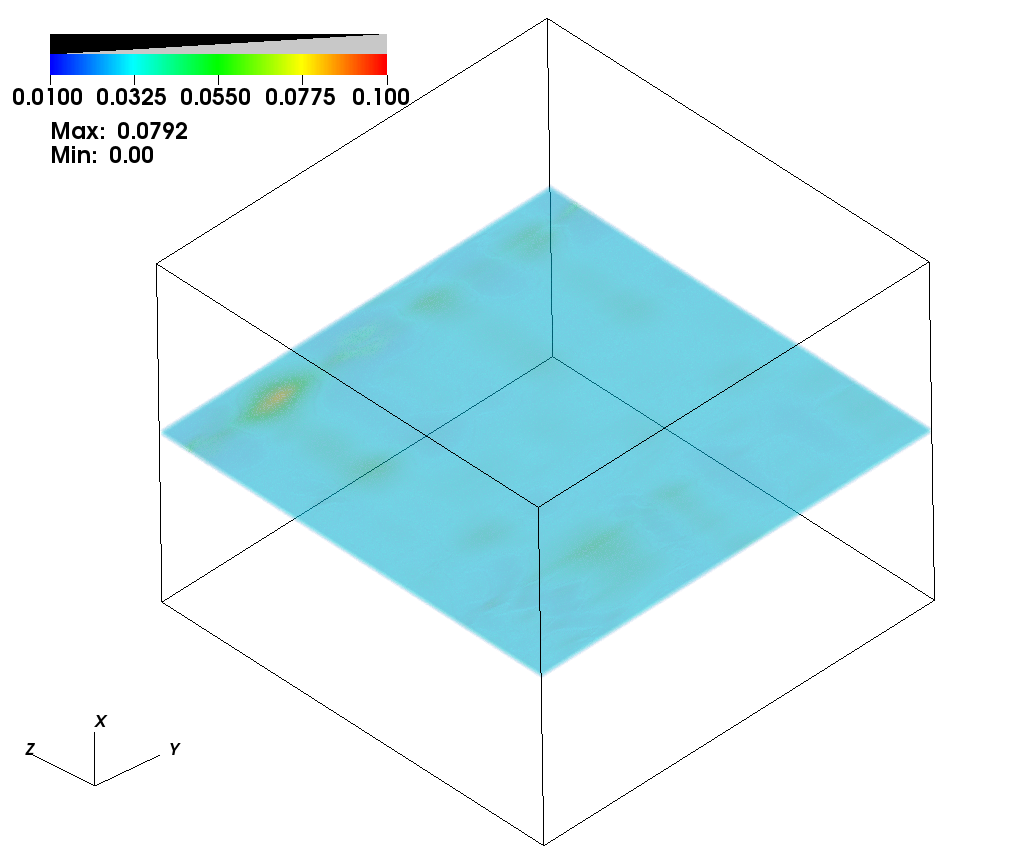}
        \subcaption{$E^{\rm res}_{\rm PB}$ at $t = 11.5 \tau_c$}\label{fig:nonid_PB_23}
    \end{minipage}

    \begin{minipage}{0.44\textwidth}
        \centering
        \includegraphics[width=\textwidth]{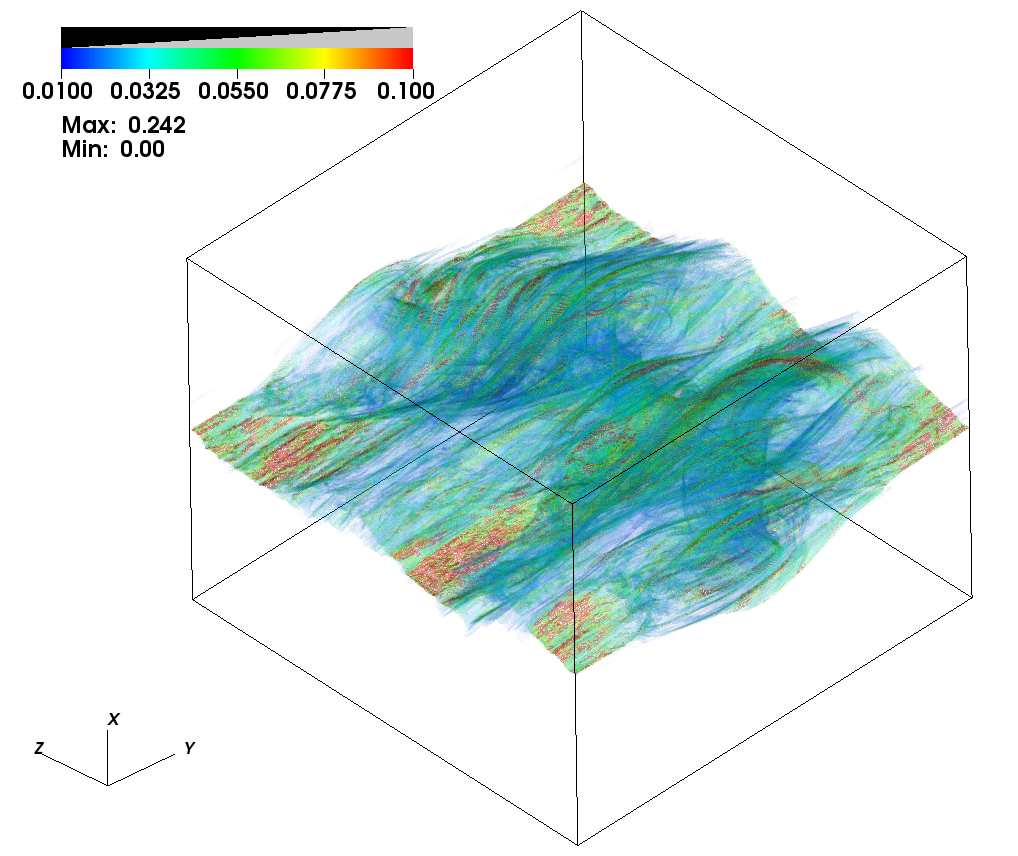}
        \subcaption{$E^{\rm res}_{\rm FF}$ at $t = 20 \tau_c$}\label{fig:nonid_FF_40}
    \end{minipage} \hspace{1cm} \vspace{5mm}
    \begin{minipage}{0.44\textwidth}
        \centering
        \includegraphics[width=\textwidth]{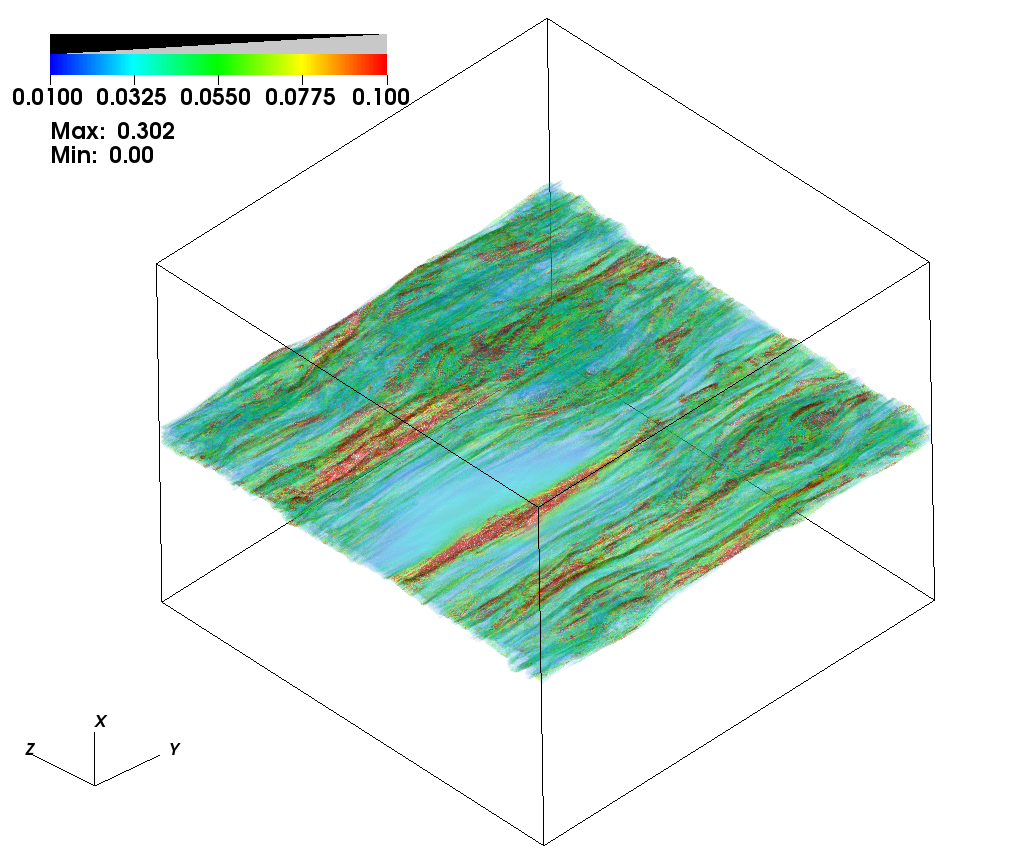}
        \subcaption{$E^{\rm res}_{\rm PB}$ at $t = 20 \tau_c$}\label{fig:nonid_PB_40}
    \end{minipage}

    \begin{minipage}{0.44\textwidth}
        \centering
        \includegraphics[width=\textwidth]{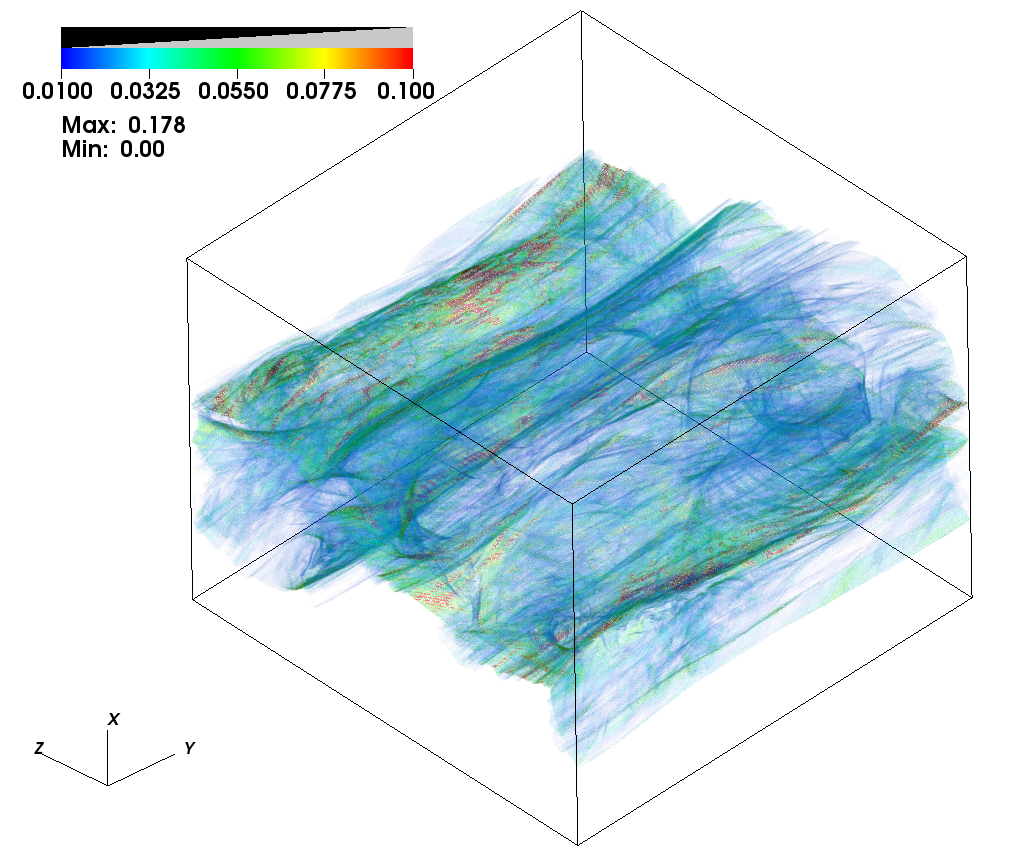}
        \subcaption{$E^{\rm res}_{\rm FF}$ at $t = 40\tau_c$}\label{fig:nonid_FF_80}
    \end{minipage} \hspace{1cm} \vspace{5mm}
    \begin{minipage}{0.44\textwidth}
        \centering
        \includegraphics[width=\textwidth]{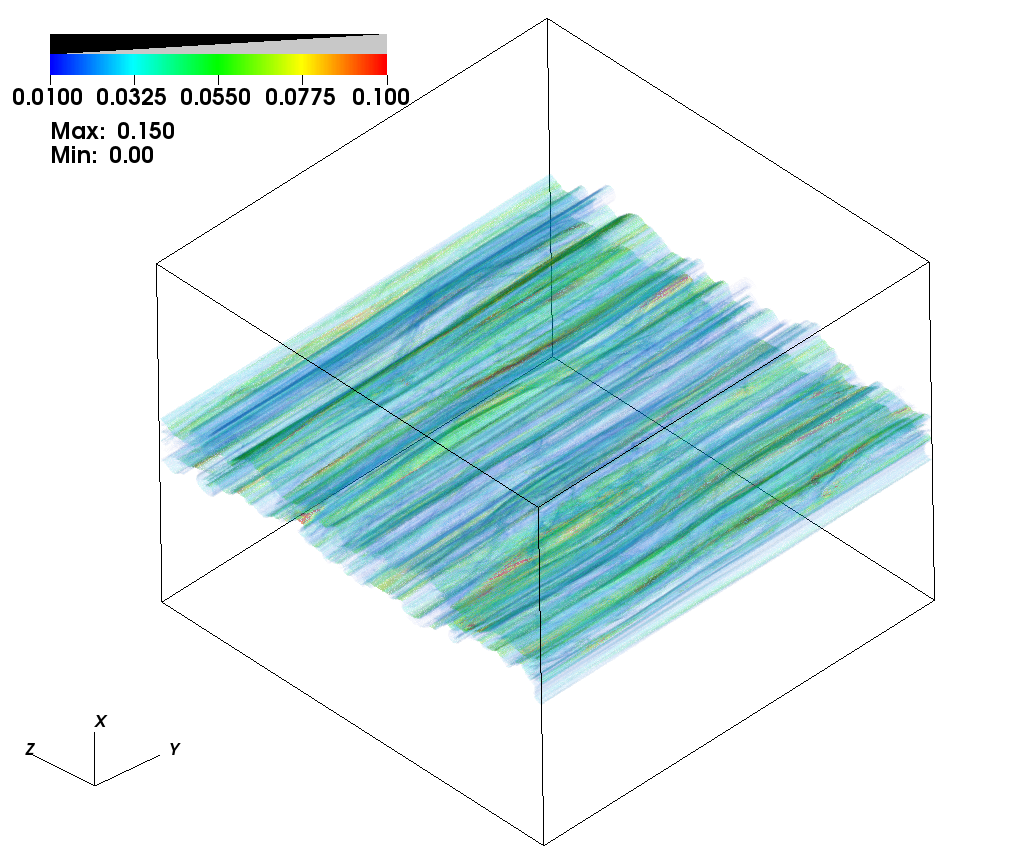}
        \subcaption{$E^{\rm res}_{\rm PB}$ at $t = 40 \tau_c$}\label{fig:nonid_PB_80}
    \end{minipage}

    \caption{\small Volume renderings of the magnitude of the resistive electric field $\vec{E}^{\rm res}=\vec{E}+\vec{v}\times\vec{B}$ for model \texttt{ff3d} ($E^{\rm res}_{\rm FF}$, left column), and for model \texttt{pb3d} ($E^{\rm res}_{\rm PB}$, right column). Snapshots are taken at times $t = 11.5\tau_c$ (Figs. \ref{fig:nonid_FF_23}, \ref{fig:nonid_PB_23}), $t = 20\tau_c$ (Figs. \ref{fig:nonid_FF_40}, \ref{fig:nonid_PB_40}), and $t = 40\tau_c$ (Figs. \ref{fig:nonid_FF_80}, \ref{fig:nonid_PB_80}).}
    \label{fig:Non_ideal}
\end{figure*}


To compare the relative efficiency of magnetic reconnection in our models we quantify their deviation from the ideal RMHD regime by focusing on the evolution of the Lorentz invariant $E^2-B^2$.
This quantity is strictly negative in the case of ideal electric fields $\vec{E}=-\vec{v}\times\vec{B}$, but it typically becomes positive in between merging plasmoids, where strong resistive electric fields are produced \citep{Sironi_2022}.

In Fig. \ref{fig:E2B2} we show on the top row the fraction of $x-$projected computational volume where $E^2>B^2$, i.e.
\begin{equation}
    \begin{aligned}
        f_{\{E^2>B^2\}}(t)
        &= \frac{1}{L_y L_z}
        \int_{0}^{L_y}
        \int_{0}^{L_z} \mathrm{d}y \mathrm{d}z \\
        &\qquad
        \Theta\!\left(
        \max_{x}
        \left[
        \mathbf{E}^2(x,y,z,t)
        -
        \mathbf{B}^2(x,y,z,t)
        \right]
        \right) \, .
    \end{aligned}
\end{equation}
Here, $\Theta$ denotes the Heaviside step function selecting only electric-dominated regions, the operator $\max_{x}$ enforces the condition that the inequality is satisfied at least at one point along each $x-$line, and the normalization by $L_y L_z = 4L^2$ yields the fraction of transverse $x-$lines at each time.
On the bottom row instead, we report the time evolution of the Lorentz invariant $E^2-B^2$ averaged over regions where it has a positive value
\begin{equation}
    \begin{aligned}
        \av{E^2-B^2}_{>0}(t)
        =
        \frac{
        \displaystyle \int_{\Omega} \mathrm{d}V\,
        \left(\mathbf{E}^2(t) - \mathbf{B}^2(t)\right)
        \; \Theta(\mathbf{E}^2 - \mathbf{B}^2)
        }
        {
        \displaystyle \int_{\Omega} \mathrm{d}V\,\;
        \Theta(\mathbf{E}^2 - \mathbf{B}^2)
        } \, ,
    \end{aligned}
\end{equation}
where $\Omega$ represents either the 2D (surface) or 3D (volume) domain where the condition $E^2>B^2$ is locally satisfied.

Both diagnostics show the quantitative impact that dimensionality and initial equilibrium configuration have on the properties of the non-ideal reconnecting regions.
The initial spike of $f_{\{E^2>B^2\}}$ (top panels of Fig. \ref{fig:E2B2}) corresponds roughly to the beginning of the non-linear phase and appears to be much higher in planar symmetry than in 3D models, regardless of the initial equilibrium configuration.
This effect could be connected to the loss of the planar symmetry in 3D models, which tends to reduce the fraction of domain hosting a considerable resistive component of the electric field. 
However, the reconnecting activity in models \texttt{ff3d} and \texttt{pb3d} appears on the other hand to last longer than the corresponding 2D runs, as the dynamics of our 3D models undergoes a much more complex saturation of the tearing instability and does not completely shut down the reconnection throughout the whole simulated period. 

The averaged Lorentz invariant measured in reconnecting regions (bottom panels of Fig. \ref{fig:E2B2}) displays similar temporal evolution in the 2D runs (dashed lines in both panels), as expected given the close resemblance of their global reconnection dynamics. 
By contrast, a clear difference emerges near the onset of the non-linear phase when considering the 3D cases (solid black lines). 
In particular, the \texttt{ff3d} model (solid line in the bottom-left panel of Fig. \ref{fig:E2B2}) develops a more pronounced peak than its 2D counterpart, indicating a stronger localized departure from ideal conditions at the beginning of the non-linear evolution.
In contrast, almost no relevant peak is visible in model \texttt{pb3d}. 
This behaviour is consistent with the suppression of plasmoid coalescence in the pressure-balanced configuration, which limits the formation of intense resistive electric fields.

The qualitative difference between our 3D runs is once again corroborated by the renderings of the magnitude of the resistive electric field $\vec{E}^{\rm res}=\vec{E}+\vec{v}\times\vec{B}$ at different times shown in Fig. \ref{fig:Non_ideal} for model \texttt{ff3d} (left column) and model \texttt{pb3d} (right column).
At $t=11.5\tau_c$, the \texttt{ff3d} model exhibits extended regions of strong non-ideal electric field distributed across the current sheet, with particularly high intensities in areas external to the flux tube forming at the time shown in the snapshot. 
At later times, the kinking of the flux tube leads to a significant deformation of the reconnecting regions outside the initial plane $x = 0$ (mid-left panel) and to the concentration of the most intense resistive electric fields within only a few localized areas of the domain. 
This diagnostic is also consistent with the formation of secondary current sheets in the vicinity of the flux tube (see also Fig. \ref{fig:FF_current_sheets}).
By the end of the simulation, the turbulent dynamic and the thickening of the reconnecting layer has significantly 
increased the extension along the $x$ direction of the reconnecting regions, with the highest values of $E^{\rm res}$ found in filamentary structures predominantly developing along the $y$ direction.

On the other hand, the non-ideal electric field present in model \texttt{pb3d} at $t=11.5\tau_c$ is much weaker and uniform than in the force-free case, due to the delayed growth of the tearing instability.
However, the system develops later on a series of filaments along the $y$ direction with a strong resistive component of the electric field, which is consistent with the general dynamics shown in Fig. \ref{fig:PB_dyn} and clearly accounts for the significant late dissipation (i.e. within the second half of the run) observed in the tridimensional version of the pressure-balanced configuration.
The distribution of the non-ideal electric field retains a similar structure up to the end of the simulation (bottom right panel), with a general weakening in magnitude, an increase in variability along $z$ (combined with a strong coherence along $y$), and a widening of the reconnecting region's thickness. 

\section{Summary and conclusions}
\label{sec:summary}
%
%

In this work we investigated relativistic magnetic reconnection triggered by the ideal tearing mode instability within the framework of resistive special-relativistic MHD.
Our study extends previous efforts that focused primarily on 2D configurations \citep[e.g.,][]{Komissarov_etal_2007, Landi_etal2015,Tenerani_2015,DelZanna_2016} by performing, for the first time, fully 3D ResRMHD simulations of ideal-tearing–driven current-sheet disruption in both force-free and pressure-balanced equilibria. 
Each setup was run in 3D domains and in corresponding 2D reference configurations to isolate the effects of dimensionality. 
The simulations were carried out with the GPU-accelerated version of the \texttt{PLUTO} code \citep{Mignone_2007, Rossazza_GPU_2025}, using the high-order numerical scheme of \citet{Mignone_Berta_2024}. 

In the force-free configuration (i.e., with a non-zero $B_z$ component), the tearing instability develops similarly in both 2D and 3D. 
In 2D, the system follows the well studied dynamics \citep[e.g.,][]{Miranda_etal2018,Mignone_etal_2019,Mignone_Berta_2024}: hierarchical plasmoid-chain reconnection drives energy release but saturates as magnetic energy remains stored in plasmoids, quenching conversion efficiency \citep{Sironi_Spitkovsky_2014, Sironi_Petropoulou_Giannios_2015}. 
In 3D, we observe the onset of ideal tearing in agreement with linear theory \citep{Pucci_Velli_2014, DelZanna_2016}, followed by parasitic plasmoid coalescence in close resemblance with the 2D reference case.
At later stages, the additional degree of freedom (the $z$-direction) enables a richer dynamics characterized by secondary instabilities (i.e., flux-rope kinking, and thickening of the layer) which excite power at progressively smaller scales along the guide-field direction. 
This process breaks the planar coherence of the layer, triggers a turbulent cascade, which fosters field dissipation. 
This interplay produces a longer-lasting reconnection and ultimately leads to greater magnetic-energy dissipation than in 2D.

The 2D pressure-balanced reconnection dynamics is qualitatively similar to that of the corresponding 2D force-free model \citep{Landi_etal2015}, though the ideal tearing instability grows at a slower rate. 
On the other hand, in the 3D pressure-balanced model with an initial null $B_z$ component, plasmoid-mediated reconnection is strongly suppressed compared to the 2D case and to force-free models. 
While the initial tearing instability develops similarly in both 2D and 3D pressure-balanced cases, the 3D model lacks the characteristic second exponential growth associated with the onset of the parasitic plasmoid coalescence instability \citep{Loureiro_2007, Loureiro_Uzdensky_2016} triggered by tearing, along with a turbulent morphology of forming plasmoids.
The forming plasmoids are disrupted by rapidly growing pressure-driven modes before they can merge, preventing the characteristic rise-and-saturation of the reconnected magnetic field profile, and the formation of a main magnetic island. 
This leads to a reduced reconnection efficiency, evidenced by weaker current spikes, and limited conversion of magnetic energy. 
A similar dynamics is observed in recent 3D first-principles closed-box simulations \citep[e.g.,][]{Werner_Uzdensky_2021}. 
Local plasma conditions on the current sheet promote the growth of pressure-driven modes, which, in the absence of a stabilizing $B_z$ component providing magnetic tension, grow rapidly along the $z-$direction and disrupt forming plasmoids. 
In 2D, such suppression cannot occur, as the lack of $z$-dependence prevents the growth of modes along $z$ and mimic the stabilizing effect of a guide field. 

To quantify how efficiently our models reconnect magnetic flux, we examined the Lorentz invariant $E^{2}-B^{2}$, which isolates regions where non-ideal electric fields become dynamically important. 
Both the fractional volume with $E^{2}>B^{2}$ and the averaged positive value of the invariant demonstrate that dimensionality and equilibrium configuration modulate the strength, morphology, and duration of reconnection. 
The initial non-linear spike in non-ideal activity is consistently higher in 2D, yet 3D models, especially the force-free case, sustain reconnection for longer owing to their more complex saturation dynamics. 

Comparison of the 3D equilibria to their respective 2D reference cases shows that the presence of a $z-$dependence in the configuration, the 3D magnetic field topology, and the local plasma conditions critically determine the dynamics and efficiency of reconnection.
The presence of a $B_z$ component in 3D enables ideal tearing to form plasmoids that persist long enough to merge, thereby triggering a bursty reconnection event.
In 2D, the lack of a $z$-dependence and a $z$-direction forces the system into planar symmetry, effectively mimicking the presence of a guide field and preventing the emergence of parasitic modes and their associated dynamics.

While 2D reconnection can be bursty due to stochastic creation, movement, and merging of plasmoids, these events tend to invariably lead toward a unique final state.
In 3D, current-sheet evolution is not simply classical reconnection with additional perturbations. 
When considering a non-zero $B_z$ component, at least three main instabilities (and beyond a single fluid description, even more, such as the Relativistic Drift Kink Instability) contribute to the dynamics: the primary one is tearing, followed by plasmoid coalescence and the flux-rope kink instability, all fueled by magnetic energy \citep{Barkov_Komissarov_o2016, Zhang_Kink_2021, Zhang_2023_kink_Flux_rope, Sironi_2025_RecReview}. 
In contrast, when $B_z$ is zero, other instabilities not necessarily powered by magnetic energy -such as pressure-driven modes- may arise and concur in affecting the development of the primary ideal-tearing event.
Thus, in 3D, the reconnection problem effectively requires the understanding of the dynamical behaviour of thin current sheets governed by the non-linear interaction of several instabilities, which may act simultaneously and either compete with or reinforce one another.
The outcome of this interplay depends sensitively on the ambient plasma, subtle configuration details, and even stochastic effects \citep{Werner_Uzdensky_2021}, and it is crucial for determining the overall evolution and efficiency of reconnection-something that can only be assessed within a fully 3D framework.

Several directions for future work emerge from our study. 
First, it is crucial to assess the sensitivity of reconnection dynamics to the initial and boundary conditions, in particular by exploring open boundary conditions and the effect of a guide field in pressure-balanced equilibria.
Secondly, recent results from particle-in-cell (PIC) simulations \citep{Selvi_etal_2023, Moran_2025} suggest that the fundamental properties governing magnetic-field dissipation in current sheets can be captured by an effective resistivity formulation. 
This approach locally enhances magnetic energy dissipation and favours the onset of fast reconnection \citep{Bugli_etal_2024, Ripperda_2026}.

Applying collisionless resistivity prescriptions to 3D ResRMHD reconnection models is therefore essential for resolving collisionless magnetic reconnection physics while achieving scale separations that are prohibitive for fully kinetic models.
Such studies are fundamental for developing reconnection models at larger scales, with potential applications including disentangling the coherent radio emission from the Crab pulsar, other young energetic pulsars, and millisecond pulsars produced in the magnetospheric current sheet beyond the light cylinder \citep{Baty_2013, Philippov_Uzdensky_Spitkovsky_Cerutti_2019}, as well as the dissipation of the striped pulsar wind in pulsar magnetospheres \citep{Cerutti_2017_Striped, Cerutti_2020_Striped}.
Moreover, coupling the findings of our simulations with large-scale models represents a crucial step toward properly isolating observable features triggered by magnetic reconnection events in simulations of black hole or neutron star magnetospheres \citep{Tchekhovskoy_2013}, AGN jets \citep{Mattia2023}, coronae \citep{Liska_2022}, accretion disks \citep{Ripperda_2022_BH_Flares}, as well as collapsing or merging objects \citep{Most_2024}.



\section*{Acknowledgements}
The authors thank B. Ripperda, L. Sironi, A. Philippov, C. Thompson, and S. Landi for useful discussions.

\vspace{7pt}\noindent
This work has received funding from the European High Performance Computing Joint Undertaking (JU) and Belgium, Czech Republic, France, Germany, Greece, Italy, Norway, and Spain under grant agreement No 101093441 (SPACE).

\vspace{7pt}\noindent
This project has received funding from the European Union's Horizon Europe research and innovation programme under the Marie Sk\l{}odowska-Curie grant agreement No. 101064953 (GR-PLUTO).

\vspace{7pt}\noindent
This paper is supported by the  Fondazione ICSC, Spoke 3 Astrophysics and Cosmos Observations. 
National Recovery and Resilience Plan (Piano Nazionale di Ripresa e Resilienza, PNRR) Project ID CN\_00000013 \quotes{Italian Research Center on High-Performance Computing, Big Data and Quantum Computing}  funded by MUR Missione 4 Componente 2 Investimento 1.4: Potenziamento strutture di ricerca e creazione di \quotes{campioni nazionali di $R\&S$ (M4C2-19 )} - Next Generation EU (NGEU).

\vspace{7pt}\noindent
We acknowledge ISCRA for awarding this project access to the Leonardo supercomputer and the EuroHPC Joint Undertaking for granting us access to the Leonardo through an EuroHPC Access call.

\vspace{7pt}\noindent
V.B. acknowledges support by a grant from the Simons Foundation (MP-SCMPS-00001470)
V.B. is in part supported by the Natural Sciences \& Engineering Research Council of Canada (NSERC) [funding reference number 568580]

\vspace{7pt}\noindent
M.B. acknowledges the support of the French Agence Nationale de la Recherche (ANR), under grant ANR-24-ERCS-0006 (project BlackJET).

\vspace{7pt}\noindent
In memory of Nuno F.G. Loureiro.

\section*{Data Availability}

The data underlying this article will be shared on reasonable request to the corresponding author.
The figures presented in this paper have been generated using the PyPLUTO package \citep{Mattia_etal_2025} and VisIt \citep{VisIt}.


\bibliographystyle{mymnras}
\bibliography{bibliography} 



\appendix

\bsp	
\label{lastpage}
\end{document}